\documentclass{aa}

\usepackage{graphicx}
\usepackage[varg]{txfonts}

\begin{document}

\title{The C$_{60}$:C$_{60}^+$ Ratio in Diffuse and Translucent Interstellar Clouds}

\author{Ga\"el Rouill\'e\inst{1}\thanks{\email{gael.rouille@uni-jena.de}}
\and Serge A. Krasnokutski\inst{1}
\and Yvain Carpentier\inst{1,2}}

\institute{Laboratory Astrophysics Group of the Max Planck Institute for Astronomy at the Friedrich Schiller University Jena, Institute of Solid State Physics, Helmholtzweg 3, D-07743 Jena, Germany
\and University of Lille, CNRS, UMR 8523 - PhLAM - Laboratoire de Physique des Lasers Atomes et Mol\'ecules, F-59000 Lille, France}

\date{Received <date> /
Accepted <date>}

\abstract{Insight into the conditions that drive the physics and chemistry in interstellar clouds is gained from determining the abundance and charge state of their components.}{We propose an evaluation of the C$_{60}$:C$_{60}^+$ ratio in diffuse and translucent interstellar clouds that exploits electronic absorption bands so as not to rely on ambiguous IR emission measurements.}{The ratio is determined by analyzing archival spectra and literature data. Information on the cation population is obtained from published characteristics of the main diffuse interstellar bands attributed to C$_{60}^+$ and absorption cross sections already reported for the vibronic bands of the cation. The population of neutral molecules is described in terms of upper limit because the relevant vibronic bands of C$_{60}$ are not brought out by observations. We revise the oscillator strengths reported for C$_{60}$ and measure the spectrum of the molecule isolated in Ne ice to complete them.}{We scale down the oscillator strengths for absorption bands of C$_{60}$ and find an upper limit of approximately 1.3 for the C$_{60}$:C$_{60}^+$ ratio.}{We conclude that the fraction of neutral molecules in the buckminsterfullerene population of diffuse and translucent interstellar clouds may be notable despite the non-detection of the expected vibronic bands. More certainty will require improved laboratory data and observations.}


\maketitle


\section{Introduction}

Mid-infrared emission observations revealed the presence of fullerene C$_{60}$ in planetary and reflection nebulae first \citep{Cami10,Sellgren10}, then in various astrophysical objects \citep[][and references therein]{Cami14}. More recently, also exploiting mid-infrared emission observations, \citet{Berne17} found that C$_{60}$ is present in the diffuse interstellar medium (ISM) as well. As to the fullerene cation C$_{60}^+$, after two decades of anticipation \citep{Foing94,Foing97}, it was identified recently as the carrier of diffuse interstellar bands (DIBs) observed at near-infrared wavelengths \citep{Campbell15,Campbell16a,Campbell16b,Walker15,Walker16,Walker17}. After several tests, the identification of C$_{60}^+$ as the carrier is considered to be firm \citep[][and references therein]{Cordiner19}, demonstrating the presence of C$_{60}^+$ in the diffuse ISM. Lastly, \citet{IglesiasGroth19} observed interstellar fullerenes C$_{60}$ and C$_{70}$, and the ions C$_{60}^+$ and C$_{60}^-$, in IR emission spectra of the star-forming region IC~348.

The ubiquity of C$_{60}$ fullerene makes it an attractive species to probe various astrophysical environments \citep[][and references therein]{Brieva16}. As an example of application, the C$_{60}$:C$_{60}^+$ ratio in the ISM gives information on the local interstellar radiation field (ISRF) and electron density since it relates to photoionization and electron recombination \citep{Bakes95}. \citet{Berne17} evaluated the C$_{60}$:C$_{60}^+$ ratio for the diffuse ISM and, from the observed mid-infrared emission bands of C$_{60}$ and the measurement of DIBs attributed to C$_{60}^+$, they obtained
\begin{quote}
		absolute abundances [...], which point to a (not very restrictive) C$_{60}$ over C$_{60}^+$ ratio ranging between 0.3 to 6.
\end{quote}
Thus the neutral species is possibly more abundant than the cation in the diffuse ISM, in agreement with a model predicting C$_{60}$:C$_{60}^+$:C$_{60}^-$ relative abundances of 0.62:0.11:0.27 (presently estimated from Fig.~5 in \citealt{Bakes95}), that is, a C$_{60}$:C$_{60}^+$ ratio of 5.6. The large uncertainty that affects the ratio given by \citeauthor{Berne17} originates chiefly in the evaluation of the column density of the neutral molecule from measurements of its IR emission spectrum. Even though improved emission observations may be a remedy, another issue arises. Indeed, the IR emission bands attributed to C$_{60}$ may actually comprise contributions by analogous species \citep{GarciaHernandez13b,Krasnokutski19}.

Electronic absorption bands are by contrast specific, also better suited for probing the diffuse ISM, hence attempts to exploit those of C$_{60}$ to evaluate the C$_{60}$:C$_{60}^+$ ratio. Resulting ratio values range from 1 \citep{Leger88} down to upper limits of 0.1 \citep{Maier17} and 0.0075 \citep{Herbig00}, the last two indicating strongly ionized populations, in contradiction with the prediction by \citet{Bakes95} and the range of values proposed by \citeauthor{Berne17}

Facing this discrepancy, we propose another determination of the C$_{60}$:C$_{60}^+$ ratio in the diffuse ISM. Information on the cation population is provided to us by published detailed measurements of DIBs that C$_{60}^+$ carries. As to the population of neutral buckminsterfullerene, it is characterized by an upper limit because the absorption bands of C$_{60}$ that are most promising as gauges, expected at wavelengths near 4000 and 6000~{\AA}, do not appear in the astronomical spectra at hand. Thus we calculate an upper limit for C$_{60}$:C$_{60}^+$ along lines of sight (LOSs) for which spectra of the 4000 and 6000~{\AA} regions are available and for which the DIBs carried by C$_{60}^+$ are documented in detail. For this purpose we evaluate anew the oscillator strength of the relevant bands in the electronic spectrum of C$_{60}$.


\section{Lines of sight}\label{sec:LOSs}

We examine seven LOSs chosen for the availability of detailed measurements of the two strongest DIBs attributed to C$_{60}^+$ \citep{Galazutdinov17} and that of spectra that cover the 4000~{\AA} and 6000~{\AA} wavelength regions where absorption bands of C$_{60}$ are expected. These LOSs exhibit a color excess $E(\mathrm{B}-\mathrm{V})$ in the range 0.2--1.48, which translates approximately into total extinction $A_\mathrm{V}$ values in the range 0.6--4.6 if we adopt the total-to-selective extinction ratio of the Milky Way \citep[$R_\mathrm{V}=3.1$,][]{Cardelli89}. Although these values suggest the LOSs cross diffuse ($A_\mathrm{V}{\la}1$) to translucent ($1{\la}A_\mathrm{V}{\la}5$) clouds, one must not presume the distribution of matter along them. Indeed, an alignment of diffuse clouds may not differ from a translucent cloud in terms of extinction. Anyhow, the spectra of all the corresponding target stars feature lines of the CH radical, indicating regions with diffuse molecular cloud conditions \citep{Snow06}. In support, all except one include lines of the CN radical. As to translucent cloud conditions, they are actually met in one case at least, toward HD~169454 \citep{Jannuzi88}.

The measurements of the 9577~{\AA} and 9632~{\AA} DIBs by \citet{Galazutdinov17} along the seven LOSs we have chosen are given in Table~\ref{tbl:C60cat}. They comprise the observed rest wavelengths $\lambda$, FWHMs $w$, and equivalent widths $W$.

Spectra of the LOS target stars measured in the 4000~{\AA} and 6000~{\AA} wavelength regions with a high signal-to-noise ratio were found in the archives of the European Southern Observatory (ESO). For each LOS, spectra measured in series in the same observational session were summed to further increase the signal-to-noise ratio. The errors on the flux values were then considered to be standard errors and were combined accordingly. The data sets are given in Appendix~\ref{apx:data}.

\begin{table*}
\caption{Column densities of C$_{60}^+$.}
\label{tbl:C60cat}
\centering
\begin{tabular}{llllll}
\hline\hline
LOS & $\lambda$ ({\AA}) & $w$ ({\AA}) & $W$ (m{\AA}) & $N$ (10$^{12}$~cm$^{-2}$) & $\Delta N$ (10$^{12}$~cm$^{-2}$) \\
\hline
HD~76341  & 9577.2 (0.2) & 2.5 (0.2) & 110 (20) & 10.5 & 6.5 \\
	        & 9632.2 (0.2) & 2.5 (0.2) & 134 (30) & 18   & 12  \\
HD~136239 & 9576.9 (0.2) & 2.9 (0.2) & 195 (15) & 19   & 11  \\
          & 9631.8 (0.1) & 2.4 (0.2) & 120 (20) & 16   & 10  \\
HD~148379 & 9577.3 (0.2) & 3.3 (0.2) & 137 (9)  & 13.2 & 7.8 \\
          & 9632.2 (0.2) & 2.4 (0.2) & 80 (11)  & 10.8 & 6.6 \\
HD~167264 & 9576.8 (0.2) & 3.2 (0.2) & 60 (17)  & 5.7  & 3.8 \\
          & 9632.4 (0.2) & 2.7 (0.2) & 82 (20)  & 11.1 & 7.2 \\
HD~168625 & 9576.2 (0.1) & 3.1 (0.2) & 320 (25) & 32   & 19  \\
          & 9631.5 (0.1) & 1.8 (0.1) & 194 (25) & 27   & 16  \\
HD~169454 & 9577.1 (0.1) & 2.3 (0.2) & 82 (10)  & 7.8\tablefootmark{a} & 4.7 \\
          & 9631.4 (0.3) & 2.7 (0.2) & 130 (20) & 18   & 11  \\
HD~183143 & 9577.3 (0.2) & 2.9 (0.2) & 300 (20) & 30\tablefootmark{a,b} & 18 \\
          & 9632.5 (0.2) & 1.9 (0.3) & 105 (20) & 14.3 & 9.0 \\
\hline
\end{tabular}
\tablefoot{The values of $\lambda$, $w$, and $W$ are from \citet{Galazutdinov17}. Values between parentheses are measurement errors.}
\tablefoottext{a}{Consistent with 9~$\times$~10$^{12}$~cm$^{-2}$ toward HD~169454 and 20~$\times$~10$^{12}$~cm$^{-2}$ toward HD~183143 in \citet{Berne17}.}
\tablefoottext{b}{Consistent with 2~$\times$~10$^{13}$~cm$^{-2}$ in \citet{Walker15} and (2 $\pm$ 0.8)~$\times$~10$^{13}$~cm$^{-2}$ in \citet{Campbell16b}.}
\end{table*}

The wavelength scales of the archival spectra are defined in the observer rest frame. For practical purpose, the standard of rest has been changed to an interstellar reference, a substance plausibly located in the same regions as interstellar fullerenes, so as to obtain scales showing the rest wavelengths of these species (Appendix~\ref{apx:shifts}).


\section{Laboratory spectra of C$_{60}$ and C$_{60}^+$}\label{sec:labspec}

\subsection{Positions and widths of the near-IR bands of C$_{60}^+$}\label{sec:labpwcat}

\citet{Campbell16b} measured the NIR spectrum of the C$_{60}^+\cdot$He complex in the gas phase below 10~K temperature. They found that absorption bands of free C$_{60}^+$ were similar in terms of wavelengths and relative intensities to those obtained for the complex, and that they matched DIBs as a consequence, leading to the first identification of one of their carriers \citep[][and references therein]{Cordiner19}. We are presently interested in the two strongest bands of C$_{60}^+$, which absorb at 9577.5 $\pm$ 0.1~{\AA} and 9632.7 $\pm$ 0.1~{\AA} according to \citet{Campbell16b}.

Exploiting the spectrum of C$_{60}^+\cdot$He further, \citeauthor{Campbell16b} determined the FWHMs of these two bands of C$_{60}^+$ for a temperature of 10~K. Denoted $w$(C$_{60}^+$, 9577.5~{\AA}) and $w$(C$_{60}^+$, 9632.7~{\AA}), they are 2.5 $\pm$ 0.2~{\AA} and 2.2 $\pm$ 0.2~{\AA}, respectively.

\subsection{Positions and widths of the near-UV and orange bands of C$_{60}$}\label{sec:labpwnt}

Measurements in a molecular beam with resonant two-photon ionization spectroscopy showed that the 1$^1$T$_{1u} \leftarrow ^1$A$_g$ transition of C$_{60}$ gives narrow absorption bands near 4000~{\AA} \citep{Haufler91}. The two strongest bands, labeled A$_0$ and A$_1$ to follow \citet{Leach92}, were thoroughly examined by \citet{Sassara01b} to assist in searches for interstellar C$_{60}$. They arise respectively at 4024 $\pm$ 0.5~{\AA} and 3980 $\pm$ 0.5~{\AA}, and, in interstellar conditions, the FWHMs $w$(C$_{60}$, A$_0$) and $w$(C$_{60}$, A$_1$) would be 4.05 $\pm$ 0.81~{\AA} and 5.54 $\pm$ 0.80~{\AA}, respectively \citep[or 25 $\pm$ 5 and 35 $\pm$ 5~cm$^{-1}$ in terms of wavenumbers in][]{Sassara01b}.

The absorption bands of C$_{60}$ near 6000~{\AA} have not been examined with the attention given those near 4000~{\AA}, possibly because they originate in electric dipole-forbidden electronic transitions \citep{Gasyna91,Leach92,Negri92,Hansen97,Orlandi02}, and hence are expected to be weaker than the near-UV bands. Ground-based astronomical measurements, however, when atmospheric absorptions are avoided, are more effective in terms of sensitivity and resolution in the 6000~{\AA} wavelength region than near 4000~{\AA}. It is thus worthwhile to examine and characterize the orange bands of C$_{60}$.

The assignment of the bands observed near 6000~{\AA} is complex because three electronic transitions are involved \citep[][and references therein]{Orlandi02}. The bands were measured over a broad wavelength domain with C$_{60}$ isolated in Ar ice at $\sim$5~K \citep{Gasyna91} and, with greater detail, in Ne ice at 4~K \citep{Sassara97} and in He droplets at 0.37~K \citep{Close97}. Measurements were also performed with C$_{60}$ in hydrocarbon solvents at room temperature and at 77~K \citep{Leach92,Hora96,Catalan02}. Prominent bands appeared to form a series and the member at longest wavelength was labeled $\gamma_0$ by \citeauthor{Leach92}

In a solvatochromism study, the $\gamma_0$ band was predicted to arise at 607.3 $\pm$ 0.2~nm in the gas phase. In another one, \citet{Renge95} observed the influence of temperature on the extrapolated gas-phase wavelength and proposed a transition energy of 16\,482 $\pm$ 24~cm$^{-1}$ for the $\gamma_0$ band of cold C$_{60}$ in the gas phase, which translates into a wavelength of 606.7 $\pm$ 0.9~nm. Finally, the spectrum of C$_{60}$ isolated in He droplets, least affected by environment-induced shifts, yielded a wavelength of 607.17~nm \citep{Close97}.

\citet{Haufler91} had already observed narrow bands near 6000~{\AA} in their study of C$_{60}$ in a molecular beam, the strongest one rising near 6070~{\AA}. \citet{Catalan94} identified this band with $\gamma_0$, the assignment of which is still uncertain \citep[][and references therein]{Orlandi02}.

We presently determine a wavelength of 6070 $\pm$ 1.0~{\AA} for the $\gamma_0$ band from the examination of the photoionization spectrum \citep[Fig.~2 in][]{Haufler91}. It is consistent with the values extrapolated using solvatochromism, the value reported for C$_{60}$ in He droplets, and particularly a later molecular-beam measurement \citep{Hansen97}. The uncertainty does not include that of the spectrometer calibration, which was not indicated.

\citeauthor{Haufler91} remarked that the FWHM of some bands was less than 5~cm$^{-1}$, which we find correct in the case of the band at 6070~{\AA}. Consequently, assuming the FWHM of the absorption band is identical to that of the measured photoionization feature, we set $w$(C$_{60}$, $\gamma_0$) to 1.5 $\pm$ 0.5~{\AA} in the cold gas phase. The temperature of the molecules in the molecular beam is uncertain, yet it is likely closer to 100~K than 10~K because the vaporization process provides the molecules with much internal energy and cooling them in a carrier gas expansion becomes less effective as their mass increases. Nevertheless, the temperature may not be higher than 100~K in the case of C$_{60}$ in a beam as estimated by \citet{Hansen97}. This is also suggested by the comparison of the FWHM presently adopted with that of simulated rotational contours \citep{Edwards93}. At 1.5 $\pm$ 0.5~{\AA}, equivalent to 4.1 $\pm$ 1.4~cm$^{-1}$, the FWHM and its uncertainty cover approximately the 50--100~K range of temperature, which corresponds to the warmer regions of a diffuse molecular cloud. In the cooler hence denser regions of diffuse clouds and in translucent ones, the FWHM would be smaller according to the simulations.

\subsection{Oscillator strengths of the near-IR bands of C$_{60}^+$}\label{sec:laboscat}

\citet{Campbell16b} derived the absorption cross sections of the bands that arise at 9577.5 and 9632.7~{\AA} in the spectrum of C$_{60}^+$. They are denoted $\sigma$(C$_{60}^+$, 9577.5~{\AA}) and $\sigma$(C$_{60}^+$, 9632.7~{\AA}) and their respective values are (5 $\pm$ 2)~$\times$~10$^{-15}$cm$^2$ and (4 $\pm$ 1.6)~$\times$~10$^{-15}$cm$^2$.

From their observational spectra, \citet{Galazutdinov17} derived FWHMs for the two strongest DIBs attributed to C$_{60}^+$, $w$(DIB, 9577~{\AA}) ranging from 2.3 $\pm$ 0.2 to 3.3 $\pm$ 0.2~{\AA}, and $w$(DIB, 9632~{\AA}) ranging from 1.8 $\pm$ 0.1 to 2.7 $\pm$ 0.2~{\AA}. The deviation toward values larger than the laboratory measurements can be attributed to a higher temperature of the interstellar ions compared with the C$_{60}^+\cdot$He complexes. It can also be attributed, at least partially, to the quality of the observational spectra. Indeed, assuming the two DIBs are caused by the same carrier, their widths should vary consistently from an LOS to another, when they do not. Moreover, deviations toward smaller values are not expected. Consequently, we assign the experimental absorption cross sections to the DIBs without considering any correction derived from their widths.

Spectra of C$_{60}^+$ isolated in Ne ice were also reported and, as expected, the bands arising at 9577.5 and 9632.7~{\AA} for hte free cation are then slightly redshifted \citep{Fulara93,Strelnikov15}. \citeauthor{Strelnikov15} attributed oscillator strengths of 0.01 $\pm$ 0.003 and 0.015 $\pm$ 0.005 to the bluer and redder bands, respectively. Thus, the relative intensities are reversed compared to gas-phase measurements, an effect one may attribute to the interaction between the cations and the Ne matrix, unless the order of the bands itself is reversed, which would be unexpected.

We can also compute effective oscillator strengths in a straightforward manner from the absorption cross sections and widths determined by \citet{Campbell16b} through
\begin{equation}\label{equ:f-s-f}
f = \frac{4\epsilon_0mc^2}{e^2}\frac{{\sigma}w_\nu}{k} ,
\end{equation}
with
\begin{equation}\label{equ:k}
k = 2\sqrt{\frac{\ln2}{\pi}} ,
\end{equation}
where $m$ and $e$ are the mass and charge of an electron, and $c$ is the speed of light in vacuum. Equation~\ref{equ:f-s-f} is obtained by making the approximation that band profiles are Gaussian in the wavenumber domain, and by using such a profile with FWHM $w_\nu$ when formulating the oscillator strength \citep[][and references therein]{Mulliken39}. In order to use $w$ from the wavelength domain as given in Sect.~\ref{sec:labpwcat}, Eq.~\ref{equ:f-s-f} can be approximated with
\begin{equation}\label{equ:f-s-l}
f = \frac{4\epsilon_0mc^2}{e^2{\lambda}^2}\frac{{\sigma}w}{k} .
\end{equation}
Values of 0.0164 $\pm$ 0.008 and 0.0114 $\pm$ 0.006 are obtained for $f$(C$_{60}^+$, 9577.5~{\AA}) and $f$(C$_{60}^+$, 9632.7~{\AA}), respectively. The reversal aside, they are consistent with those derived by \citeauthor{Strelnikov15} We retain the strengths derived from the absorption cross sections measured by \citet{Campbell16b} on the C$_{60}^+\cdot$He complex because the spectrum of C$_{60}^+$ is less affected by the interaction with a single He atom than by the interaction with several Ne atoms, and because the gas-phase spectrum of the C$_{60}^+\cdot$He complex is in better agreement with the DIB spectrum in terms of relative strengths compared to the matrix-isolation measurements.

\subsection{Oscillator strengths of the near-UV and orange bands of C$_{60}$}\label{sec:labosnt}

\citet{Haufler91} determined an oscillator strength of 0.006 $\pm$ 0.002 for the peaks of the 1$^1$T$_{1u} \leftarrow ^1$A$_g$ transition, that is, \citeauthor{Leach92}'s system of A$_n$ bands. They did so by comparing their spectrum of C$_{60}$ in a molecular beam with a spectrum obtained from a solution. Yet, from the spectrum of C$_{60}$ in hexane at room temperature, \citet{Leach92} obtained a value of 0.015 $\pm$ 0.005 for the same transition. The discrepancy between the two values possibly originates in the fact that \citeauthor{Leach92} did not correct the oscillator strength for the effect of the solvent \citep{Chupka97}. Considering a refractive index of 1.39 $\pm$ 0.01 for hexane at room temperature at 400~nm \citep[estimated from Fig. 3 in][]{Sowers72}, a correction factor of 0.582 $\pm$ 0.008 is obtained, which leads to an oscillator strength of 0.0087 $\pm$ 0.0031 for the 1$^1$T$_{1u} \leftarrow ^1$A$_g$ transition of free C$_{60}$. This value is in good agreement with that reported by \citeauthor{Haufler91}, which is consequently chosen for our study.

We have determined the oscillator strengths of the A$_0$ and A$_1$ bands by distributing the value given by \citeauthor{Haufler91} among the narrow peaks measured in a spectrum of C$_{60}$ isolated in Ne ice, in proportion to their areas, which we assume to be similarly affected by the matrix environment. The spectroscopy is briefly described in Appendix~\ref{apx:MIS}. Figure~\ref{fig:MIS} shows the matrix-isolation spectrum and Gaussian profiles fitted onto seven peaks of the 1$^1$T$_{1u} \leftarrow ^1$A$_g$ transition, that is, the system of A$_n$ bands. The fitting procedure gives for the seven peaks together an area of 6.64 $\pm$ 0.21~cm$^{-1}$, with individual peak areas of 1.386 $\pm$ 0.064~cm$^{-1}$ for A$_0$ and 2.938 $\pm$ 0.073~cm$^{-1}$ for A$_1$. Thus, values of 0.00131 $\pm$ 0.00056 and 0.0027 $\pm$ 0.0011 are derived for $f$(C$_{60}$, A$_0$) and $f$(C$_{60}$, A$_1$), respectively. They are visibly lower than those used by \citeauthor{Berne17}, 0.005 and 0.007, yet on the same order.

\begin{figure}
\resizebox{8.8cm}{!}{\includegraphics{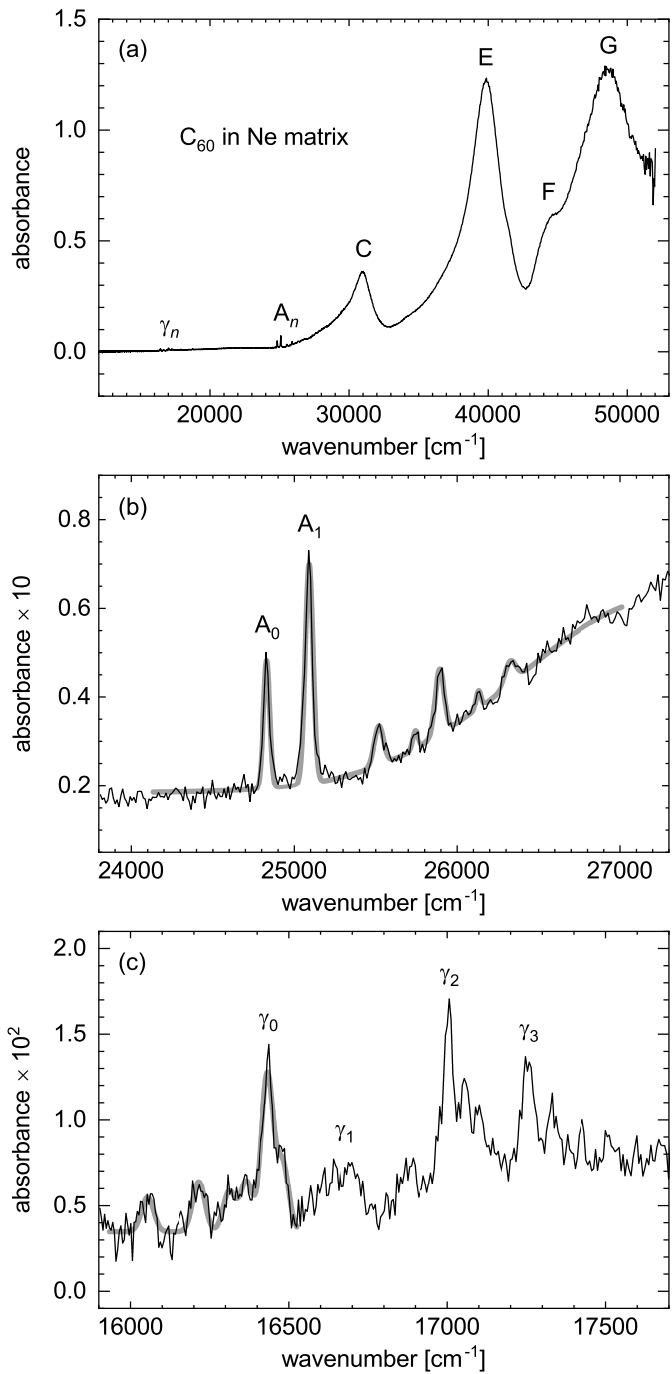}}
\caption{(a) Absorption spectrum of C$_{60}$ isolated in a Ne matrix, not corrected for scattering. Band labels according to \citet{Leach92}. (b) Gaussian profiles (thick gray solid curve) fitted to bands of the 1$^1$T$_{1u} \leftarrow ^1$A$_g$ transition, including A$_0$ and A$_1$, and to the baseline. (c) Gaussian profiles (thick gray solid curve) fitted, with a common FWHM, to absorptions near the $\gamma_0$ band. Fitted band positions are 16\,431.1 $\pm$ 1.2 cm$^{-1}$, 24\,829.04 $\pm$ 0.98 cm$^{-1}$, and 25\,087.73 $\pm$ 0.65 cm$^{-1}$ for $\gamma_0$, A$_0$, and A$_1$, respectively. The wavelength scale of the spectrometer was calibrated with an accuracy of 0.3~nm.}
\label{fig:MIS}
\end{figure}

Similarly, the oscillator strength of the orange band $\gamma_0$ is obtained by comparing its area of 0.435 $\pm$ 0.022~cm$^{-1}$ with that of the A$_1$ band (Fig.~\ref{fig:MIS}). Consequently $f$(C$_{60}$, $\gamma_0$) equals 0.00040 $\pm$ 0.00016.


\section{Astronomical spectra}\label{sec:obsspec}

\subsection{Observational data near 4000~{\AA}}\label{sec:obs4000}

Several atomic lines arise in stellar spectra around 4000~{\AA}. The strongest lines are the \ion{H}{I} (H$\epsilon$) and \ion{He}{I} lines at 3970.075 and 4026.1914~{\AA} rest wavelengths, respectively. While the \ion{H}{I} and \ion{He}{I} are observed in all spectra, the other lines appear with relative strengths that vary widely depending on the target star.

\begin{figure}
\resizebox{8.8cm}{!}{\includegraphics{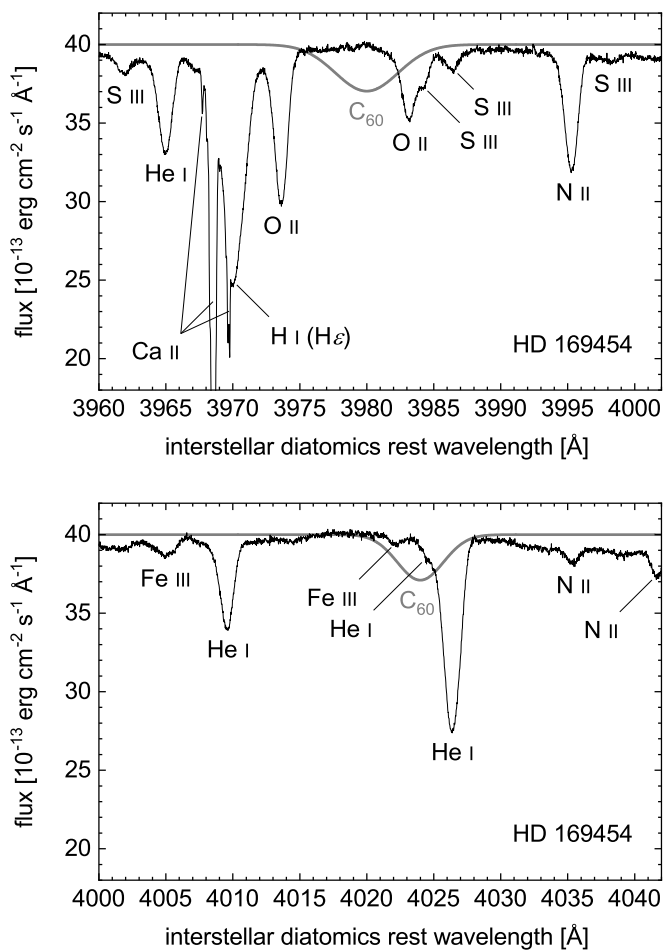}}
\caption{Spectrum toward HD~169454 (black solid curve) and synthetic absorption spectrum of C$_{60}$ with the expected band positions, widths, and relative areas (thick gray solid curve). Narrow absorption lines in the 3967--3971~{\AA} interval correspond to velocity components of the \ion{Ca}{II} H line.}
\label{fig:lines}
\end{figure}

In each observational spectrum taken into account, the A$_0$ band of C$_{60}$ expected at 4024~{\AA} would overlap the strong \ion{He}{I} line mentioned above. Figure~\ref{fig:lines} illustrates the phenomenon in the spectrum of HD~169454; it also shows that the A$_0$ band and a minor \ion{He}{I} line (4023.973~{\AA} rest wavelength) are superimposed, not to mention the overlap by an \ion{Fe}{III} line \citep[4022.35~{\AA} rest wavelength,][and references therein]{Thompson08}.

\begin{figure}
\resizebox{8.8cm}{!}{\includegraphics{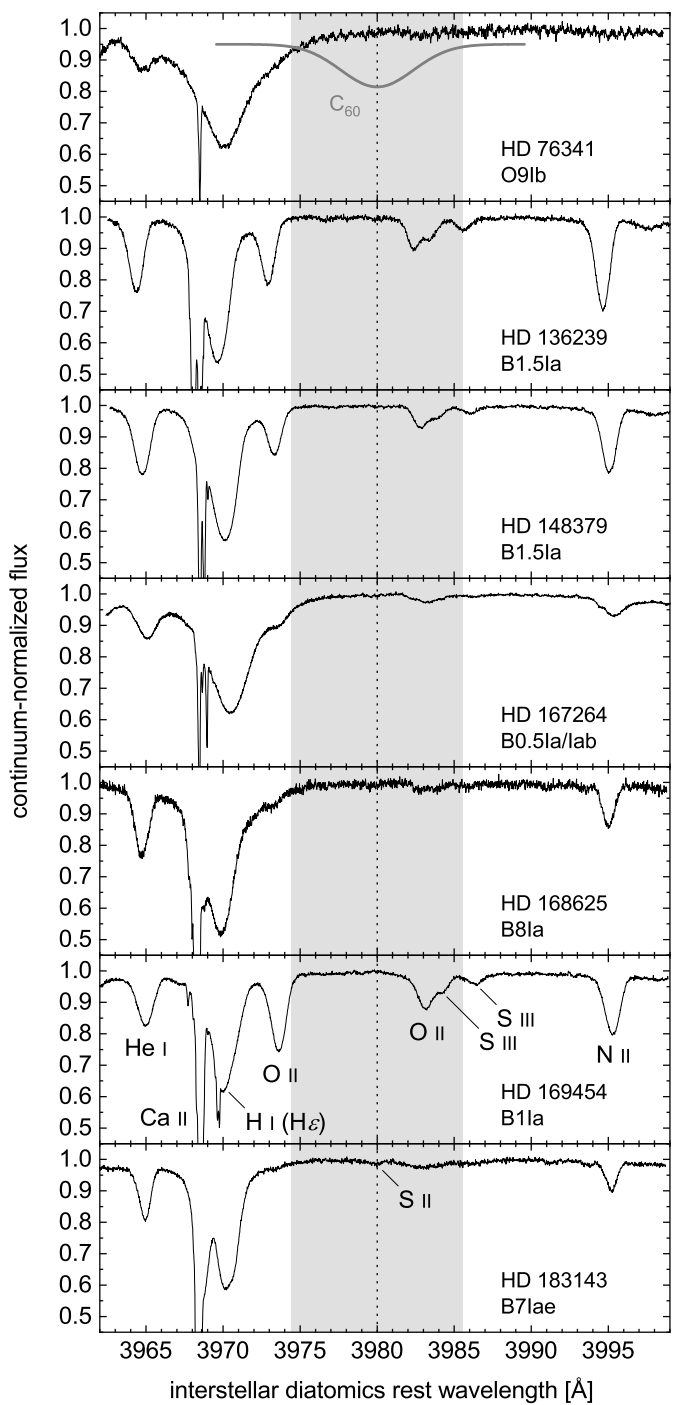}}
\caption{Normalized observational spectra (black solid curves). A straight line continuum was defined over the 3960--4000~{\AA} range of each spectrum still in the observer rest frame. The top panel includes a synthetic A$_1$ band of C$_{60}$ with expected position and width, and arbitrary equivalent width (thick gray solid curve). Its position is indicated through all panels (vertical dotted line). The light gray area indicates the wavelength interval of two FWHMs taken into account to evaluate $^{\mathrm{u}}W$(C$_{60}$, A$_1$) at 3980~{\AA} for each LOS. The spectral type of each target star is indicated.}
\label{fig:s4000A}
\end{figure}

In comparison, the A$_1$ band of C$_{60}$ at 3980~{\AA} is better separated from the nearby \ion{O}{II} line found at 3982.7140~{\AA} rest wavelength, and from \ion{S}{III} lines that rise at longer wavelengths. It is superimposed with a line of \ion{S}{II} (3979.829~{\AA} rest wavelength), which is apparent toward HD~183143 as observed in Fig.~\ref{fig:s4000A}. Figure~\ref{fig:s4000A} also shows that the strength of the \ion{O}{II}, \ion{S}{II}, and \ion{S}{III} lines depends on the spectral type of the target star and can be extremely weak.

Overlap with stellar lines is critical when the interstellar band is comparatively much weaker or, a fortiori, not observed because it lowers the signal-to-noise ratio that is taken into account to determine the column density of the interstellar species or its upper limit. In such instance, concerning the 4000~{\AA} region, the A$_1$ band of C$_{60}$ at 3980~{\AA} is preferred to A$_0$, although broadening of the H$\epsilon$ line such as toward O9Ib-type HD~76341 may constitute a cause for concern.

\subsection{Observational data near 6000~{\AA}}\label{sec:obs6000}

\begin{figure}
\resizebox{8.8cm}{!}{\includegraphics{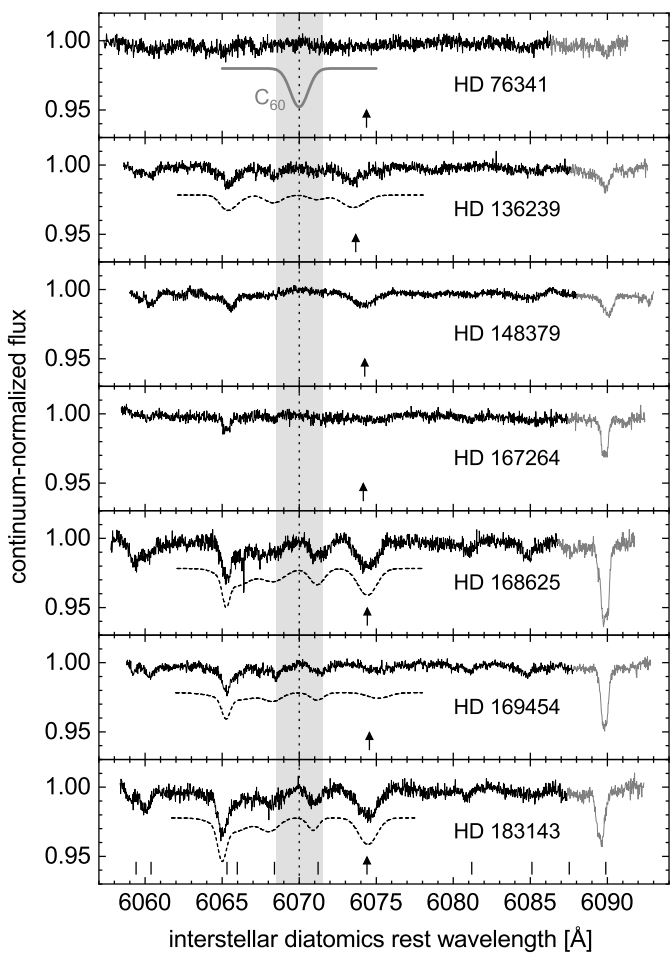}}
\caption{Continuum-normalized spectra with interstellar wavelength scale for seven LOS in the 6000~{\AA} region. Each gray portion was obtained by extrapolating the straight continuum defined over the 6058--6087~{\AA} interval (observer rest frame). The top panel includes a synthetic $\gamma_0$ band of C$_{60}$ of arbitrary equivalent width (thick gray solid curve, offset for clarity). Its position is indicated through all panels (vertical dotted line). Positions of DIBs according to \citet{Hobbs09} are indicated in the bottom panel (short vertical lines). The predicted position of a line of stellar \ion{Ne}{I} is marked in each panel (arrow). Synthetic spectra of DIBs and the \ion{Ne}{I} line are represented in four panels (short-dashed curve). The light gray area indicates the wavelength interval of two FWHMs taken into account to evaluate $^{\mathrm{u}}W$(C$_{60}$, $\gamma_0$) at 6070~{\AA} for each LOS.}
\label{fig:s6000A}
\end{figure}

Figure~\ref{fig:s6000A} shows the spectral region where the $\gamma_0$ band of C$_{60}$ is expected. A straight line continuum was defined over the 6058--6087~{\AA} range of each original spectrum, still in the observer rest frame, and used to normalize the corresponding spectrum. The straight continuum was extrapolated to 6092~{\AA} so as to be applied to the 6090~{\AA} narrow DIB, the position of which was to serve as reference for evaluating those of other DIBs (Sect.~\ref{sec:Nnt}). Next the wavelength scale was changed from observer rest frame to interstellar (Appendix~\ref{apx:shifts}).

While the 4000~{\AA} spectral region is rich in stellar lines, the 6000~{\AA} region comprises several DIBs and, in most cases, a line of stellar \ion{Ne}{I} measured at 6074.338~{\AA} in the laboratory \citep[see][for HD~183143]{Hobbs09}. For all LOSs but HD~169454, the position of this line has been predicted from the position of the \ion{He}{I} line measured at 5875.62~{\AA} in the laboratory. As to the spectrum toward HD~169454, we used lines of \ion{N}{II}, \ion{Al}{III}, and \ion{Si}{III} that are observed clearly between 5665 and 5740~{\AA}, for the \ion{He}{I} line does not have a peak profile that would yield an accurate position.

A shift is apparent between the DIBs in the spectrum toward HD~183143 and the positions reported by \citet{Hobbs09} for this very LOS (Fig.~\ref{fig:s6000A}). It is caused by a difference in the choice of reference for rest, the strongest velocity component of diatomic lines versus the strongest component of the 7698~{\AA} \ion{K}{I} line. The velocity difference is $\sim$15~km~s$^{-1}$ \citep{Hobbs09}, representing 0.3~{\AA} at 6000~{\AA} as observed in Fig.~\ref{fig:s6000A}.

Finally, although H$_2$O vapor gives numerous absorption lines near 6070~{\AA} \citep{Carleer99}, they are very weak and generally not detected. Thus telluric features are not seen.

Two DIBs are found at 6068.4 and 6071.1~{\AA} \citep{Galazutdinov00b,Tuairisg00,Weselak00,Hobbs09}, close to the position of the $\gamma_0$ band, that is, 6070 $\pm$ 1~{\AA}. Their FWHMs are on the order of 1--2~{\AA}, as expected for the $\gamma_0$ band. There is however no further indication at this stage that either of these DIBs could be identified with the latter. Although they represent rather weak absorptions, their presence influences the search for the C$_{60}$ feature.


\section{Column density of C$_{60}^+$}\label{sec:Ncat}

The column density of C$_{60}^+$ can be determined by comparing the equivalent widths of the relevant DIBs with Gaussian profiles defined using the absorption cross sections and associated widths derived by \citet{Campbell16b} for a temperature of 10~K. We assume here that, for a given column density, $W$ is conserved over a broad range of conditions. Considering an absorption feature with a Gaussian profile in the wavelength domain, where $W$ is defined, the column density $N$ of the absorbing species derived from the Lambert-Beer law is such that
\begin{equation}\label{equ:N-Gauss}
N = -\frac{1}{\sigma}\ln\left[1-k\frac{W}{w}\right] .
\end{equation}
Given the large uncertainties on $\sigma$(C$_{60}^+$, 9577.5~{\AA}) and $\sigma$(C$_{60}^+$, 9632.7~{\AA}), both 40\%, on $w$(C$_{60}^+$, 9577.5~{\AA}) and $w$(C$_{60}^+$, 9632.7~{\AA}), respectively 8\% and 9\%, and on $W$, 6--16\%, $\Delta N$ is computed by using the propagation formula for large errors proposed by \citet{Seiler87}. We assume here that the errors are independent and correspond to normal distributions. A correction is applied accordingly to $N$ as calculated with Eq.~\ref{equ:N-Gauss} because the right hand term is a function that does not conserve the symmetry of normal distributions \citep{Seiler87}. The equivalent widths $W$ of the DIBs observed at $\sim$9577 and $\sim$9632~{\AA}, the column densities $N$(C$_{60}^+$) derived from them, and the corresponding errors, are given in Table~\ref{tbl:C60cat}.

In several cases, $N$(C$_{60}^+$, 9577.5~{\AA}) and $N$(C$_{60}^+$, 9632.7~{\AA}) differ by a factor $\sim$2 although they are derived for the same LOS. Taking the uncertainties into account, the two values in each pair are nonetheless consistent. We note that the values presently obtained for $N$(C$_{60}^+$, 9577.5~{\AA}) toward HD~169454 and HD~183143, respectively (7.8 $\pm$ 4.7)~$\times$~10$^{12}$ and (30 $\pm$ 18)~$\times$~10$^{12}$~cm$^{-2}$, are in agreement with those derived by \citeauthor{Berne17}, respectively 9~$\times$~10$^{12}$ and 20~$\times$~10$^{12}$~cm$^{-2}$ (Table~C.1 in \citealt{Berne17}).


\section{Column density of C$_{60}$}\label{sec:Nnt}

Bands of interstellar C$_{60}$ have been looked for near 4000~{\AA} without success \citep[e.g.,][]{Herbig00,Sassara01b,GarciaHernandez12,GarciaHernandez13a,DiazLuis15}. Similarly, the seven spectra in Fig.~\ref{fig:s4000A} do not show any clear indication of an absorption band around 3980.0~{\AA} that could be attributed to C$_{60}$, and Fig.~\ref{fig:s6000A} does not reveal any new feature close to 6070.0~{\AA} either. As a consequence, column densities $N$ cannot be obtained for neutral buckminsterfullerene. Nevertheless, upper limits denoted $^{\mathrm{u}}N$ can be estimated.

Values of $^{\mathrm{u}}N$ can be evaluated by combining Eqs.~\ref{equ:f-s-l} and \ref{equ:N-Gauss}. Observing that $kW$ is small compared to $w$, and replacing $N$ and $W$ by the corresponding upper limits $^{\mathrm{u}}N$ and $^{\mathrm{u}}W$, we obtain
\begin{equation}\label{equ:uN}
^{\text{u}}N = \frac{4 \epsilon_0 m c^2}{e^2 \lambda^2} \frac{^{\text{u}}W}{f} ,
\end{equation}
or
\begin{equation}\label{equ:uN-num}
^{\text{u}}N = 1.1296 \times 10^{17} \frac{^{\text{u}}W}{\lambda^2 f} ,
\end{equation}
with $^{\mathrm{u}}N$, $^{\mathrm{u}}W$, and $\lambda$ expressed in cm$^{-2}$, m{\AA}, and {\AA}, respectively. One recognizes the formula used by \citet{Herbig00}, \citet{DiazLuis15}, or \citet{Berne17}. We calculate $\Delta ^{\text{u}}N$ and correct $^{\text{u}}N$ following \citeauthor{Seiler87} to take the large error on $f$ into account.

\begin{figure}
\resizebox{8.8cm}{!}{\includegraphics{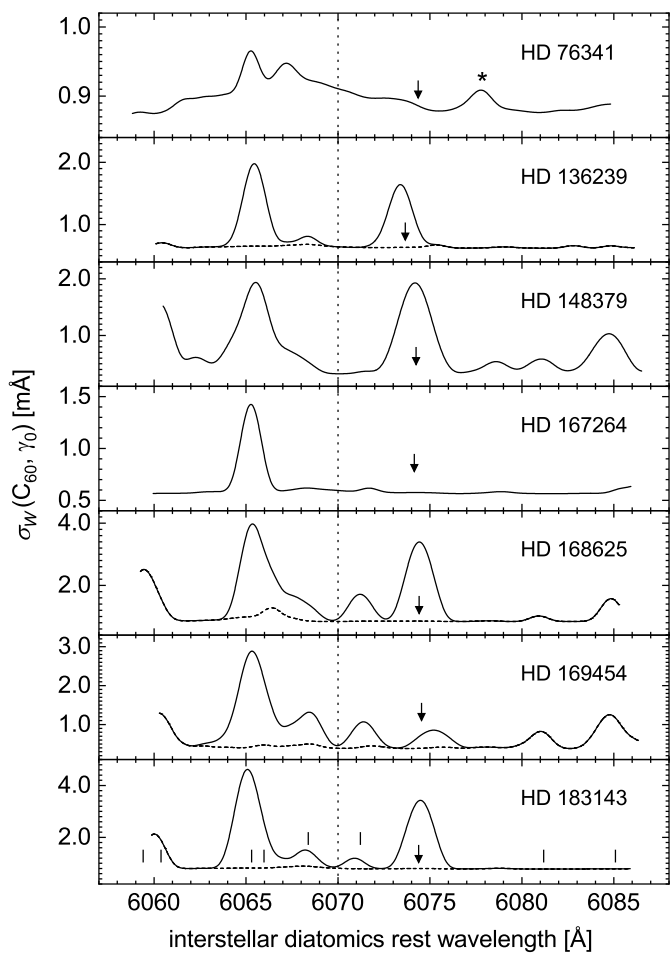}}
\caption{Equivalent-width detection threshold in the 6000~{\AA} region for seven LOS (solid curves). The vertical dotted line shows the expected position of the $\gamma_0$ band of C$_{60}$ in all spectra. Positions of DIBs according to \citet{Hobbs09} are indicated in the bottom panel (short vertical lines). The predicted position of a line of stellar \ion{Ne}{1} is marked in each panel (arrow). In four cases detection limits were computed as fitted profiles of the DIBs nearest 6070~{\AA} were included in the continuum (short-dashed curves). An unidentified absorption would cause the rise near 6078~{\AA} (asterisk).}
\label{fig:lim6000A}
\end{figure}

Values of $^{\mathrm{u}}W$, the upper limit for the equivalent width of the band expected at wavelength $\lambda$, are taken equal to three equivalent-width detection thresholds, that is, $3\sigma_W$, where $\sigma_W$ is computed according to \citet{Lawton08}. Figure~\ref{fig:lim6000A} presents $\sigma_W$ curves for the 6000~{\AA} region. Because $\sigma_W$ depends on the difference between the observed spectrum and the chosen continuum, $\sigma_W$ peaks at the position of any band or line that has not been included in the continuum, as confirmed by a comparison with Fig.~\ref{fig:s6000A}. Interestingly, the $\sigma_W$ curves reveal a possible absorption feature at a position varying from 6078 to 6079~{\AA}, which we have not identified. The varying wavelength suggests it is not a DIB, though, unless it is the effect of a poorly defined shape.

Moreover, the value of $\sigma_W$ at a given wavelength considers the difference between the observed spectrum and the chosen continuum for all wavelengths in a  surrounding interval spanning two FWHMs of the band to be detected. The value of $\sigma_W$ at this wavelength can then be lowered by including in the continuum any nearby band with a profile that extends into the interval. Figure~\ref{fig:lim6000A} includes cases in which fitted profiles of the DIBs nearest 6070~{\AA} were included in the continuum. In the fitting procedure, each DIB was given a Gaussian profile with a position fixed to literature value \citep{Hobbs09} -- corrected using the shift observed for the 6090~{\AA} DIB -- while its FWHM and area were free. A poor fit results in an incomplete elimination of the feature contribution to the $\sigma_W$ curve.

\begin{figure}
\resizebox{8.8cm}{!}{\includegraphics{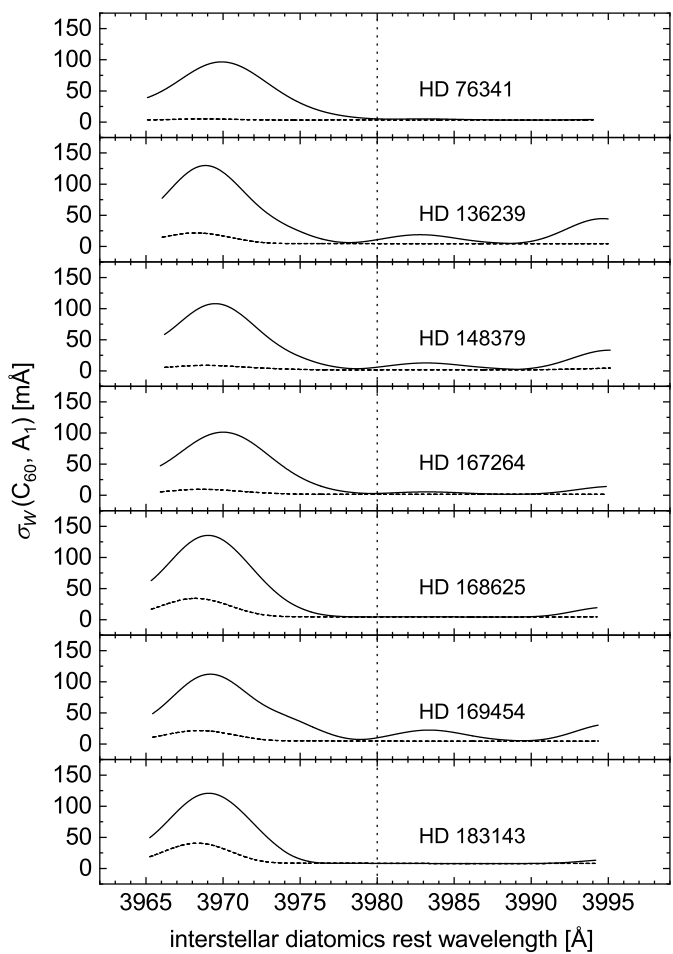}}
\caption{Equivalent-width detection threshold in the 4000~{\AA} region scale for seven LOS (solid curves). The vertical dotted line shows the expected position of the near-UV band of C$_{60}$ in all spectra. The detection limits were also computed as fitted profiles of stellar lines were included in the continuum (short-dashed curves).}
\label{fig:lim4000A}
\end{figure}

Figure~\ref{fig:lim4000A} shows $\sigma_W$ curves for the 4000~{\AA} region. They are given both with the stellar and interstellar atomic lines kept in the spectra and also with fitted profiles of the lines included in the continua. Large residuals from fitting the overlapping multiple components of the \ion{Ca}{II} line cause the broad peak at 3968.5~{\AA} that appears after the procedure.

Table~\ref{tbl:C60} presents values for $^{\mathrm{u}}W$(C$_{60}$) and $^{\mathrm{u}}N$(C$_{60}$) obtained from the normalized observational spectra shown in Figs.~\ref{fig:s4000A} and \ref{fig:s6000A}. They were derived by giving $f$(C$_{60}$, A$_1$), $w$(C$_{60}$, A$_1$), $f$(C$_{60}$, $\gamma_0$), $w$(C$_{60}$, $\gamma_0$), and their uncertainties the values introduced and adopted in Sects.~\ref{sec:labpwnt} and \ref{sec:labosnt}. The equivalent-width detection thresholds $\sigma_W$ yielding the $^{\mathrm{u}}W$ values were computed by using the data comprised within an interval centered at the rest wavelength of the relevant band and spanning two FWHMs.

\begin{table*}
\caption{Upper limits for column densities of interstellar C$_{60}$ considering the A$_1$ and $\gamma_0$ bands.}
\label{tbl:C60}
\centering
\begin{tabular}{lllllllll}
\hline\hline
LOS & \multicolumn{4}{c}{C$_{60}$, A$_1$ (3980~{\AA})} & \multicolumn{4}{c}{C$_{60}$, $\gamma_0$ (6070~{\AA})} \\
 & $^{\mathrm{u}}W$ & $\Delta ^{\mathrm{u}}W$\tablefootmark{a} & $^{\mathrm{u}}N$ & $\Delta ^{\mathrm{u}}N$ & $^{\mathrm{u}}W$ & $\Delta ^{\mathrm{u}}W$\tablefootmark{a} & $^{\mathrm{u}}N$ & $\Delta ^{\mathrm{u}}N$ \\
 & (m{\AA}) & (m{\AA}) & (10$^{12}$~cm$^{-2}$) & (10$^{12}$~cm$^{-2}$) & (m{\AA}) & (m{\AA}) & (10$^{12}$~cm$^{-2}$) & (10$^{12}$~cm$^{-2}$) \\
\hline
HD~76341  & 15.3          & 5.3           & 50            & 36           & 2.7           & 1.1           & 26          & 19          \\
					& \textit{10.3} & \textit{2.0}  & \textit{34}   & \textit{22}  & ...           & ...           & ...         & ...         \\
HD~136239 & 32.1          & 7.9           & 106           & 70           & 1.93          & 0.79          & 18          & 13          \\
					& \textit{13.1} & \textit{2.3}  & \textit{43}   & \textit{27}  & \textit{1.93} & \textit{0.79} & \textit{18} & \textit{13} \\
HD~148379 & 17.5          & 4.6           & 58            & 38           & 0.97          & 0.39          & 9.2         & 6.7         \\
					& \textit{4.57} & \textit{0.78} & \textit{15.1} & \textit{9.5} & ...           & ...           & ...         & ...         \\
HD~167264 &  8.4          & 2.8           & 28            & 19           & 1.78          & 0.73          & 17          & 12          \\
					& \textit{4.25} & \textit{0.75} & \textit{14.0} & \textit{8.9} & ...           & ...           & ...         & ...         \\
HD~168625 & 13.1          & 2.2           & 43            & 27           & 2.8           & 1.1           & 27          & 19          \\
					& \textit{13.1} & \textit{2.2}  & \textit{43}   & \textit{27}  & \textit{2.5}  & \textit{1.0}  & \textit{24} & \textit{17} \\
HD~169454 & 29.6          & 6.8           & 98            & 64           & 1.37          & 0.59          & 13.0        & 9.7         \\
					& \textit{14.1} & \textit{2.5}  & \textit{46}   & \textit{29}  & \textit{1.16} & \textit{0.46} & \textit{11.0} & \textit{8.0} \\
HD~183143 & 24.5          & 4.3           & 81            & 51           & 2.7           & 1.0           & 26          & 18          \\
					& \textit{24.5} & \textit{4.3}  & \textit{81}   & \textit{51}  & \textit{2.41} & \textit{0.97} & \textit{23} & \textit{17} \\
\hline
\end{tabular}
\tablefoot{Values in italic were obtained by including in the continuum the fitted profiles of atomic stellar lines when in the 4000~{\AA} region, and the four DIBs nearest 6070~{\AA} and that of the nearby \ion{Ne}{I} line when in the 6000~{\AA} region.}
\tablefoottext{a}{Uncertainty computed according to \citet{Lawton08}.}
\end{table*}

According to the contents of Table~\ref{tbl:C60}, incorporation of specific spectral features into the continuum has the potential to lower substantially the upper limit for equivalent width. The most spectacular improvement is seen with HD~148379 where the $^{\mathrm{u}}W$ and $^{\mathrm{u}}N$ values derived from the UV spectrum are divided by a factor close to four. The limits $^{\mathrm{u}}N$(C$_{60}$, A$_1$) and $^{\mathrm{u}}N$(C$_{60}$, $\gamma_0$) presented in Table~\ref{tbl:C60} are in the ranges 14--81~$\times$~10$^{12}$~cm$^{-2}$ and 9--26~$\times$~10$^{12}$~cm$^{-2}$, respectively, when only the values obtained by including spectral features in the continuum, whenever available, are considered. The corresponding averages are 39~$\times$~10$^{12}$~cm$^{-2}$ and 18~$\times$~10$^{12}$~cm$^{-2}$. Thus $^{\mathrm{u}}N$(C$_{60}$, A$_1$) is approximately twice as large as $^{\mathrm{u}}N$(C$_{60}$, $\gamma_0$). The uncertainties are very large, with 63--65\% for $^{\mathrm{u}}N$(C$_{60}$, A$_1$) and 70--74\% for $^{\mathrm{u}}N$(C$_{60}$, $\gamma_0$).


\section{Upper limits for the C$_{60}$:C$_{60}^+$ ratio}\label{sec:ul-nt:cat}

In order to compare populations of C$_{60}$ and C$_{60}^+$ in interstellar clouds, we assume that the angular size of the clouds before a target star is large in comparison with the difference in LOS coordinates during the observations at UV and visible wavelengths, which remains smaller than 5" (Table~\ref{apx:data}). We also assume that the properties of the clouds do not vary significantly between the two sets of coordinates.

Our comparison is relevant to entire clouds and not to specific layers because the wavelengths of the DIBs attributed to C$_{60}^+$ were determined by calibrating astronomical spectra with both atomic and molecular lines \citep{Galazutdinov17}. The DIB wavelengths are given with an error in the range 0.1--0.3~{\AA}, that is, on the order of the equivalent velocity difference between the diffuse atomic outer layer and the denser diatomic layer observed for HD~183143 (see Sect.~\ref{sec:obs6000}). Moreover, the bands of C$_{60}$ are currently not characterized with sufficient precision to allow us to search for them in a given layer.

The upper limit of the C$_{60}$:C$_{60}^+$ ratio then equals $^{\mathrm{u}}N$(C$_{60}$)/$N$(C$_{60}^+$). It is calculated by combining Eqs.~\ref{equ:N-Gauss} and \ref{equ:uN-num} into which the values given in Tables~\ref{tbl:C60cat} and \ref{tbl:C60} are entered, and by adding a correction that large errors make necessary \citep{Seiler87}. The errors that affect the ratios are then obtained by taking into account those attached to the band descriptors rather than $\Delta ^{\mathrm{u}}N$ and $\Delta N$. The results are given in Table~\ref{tbl:C60:C60cat}, showing that the upper limit of the C$_{60}$:C$_{60}^+$ ratio noticeably depends on the band used to evaluate $^{\mathrm{u}}N$(C$_{60}$). It ranges from 1.6 to 6.9 with an unweighted average value of 3.5 when using A$_1$, from 0.90 to 5.2 averaging 1.8 when using $\gamma_0$. While the values in the latter case are lower and less spread than in the former when the extreme value of 5.2 is excluded, their relative uncertainties are greater. In either case the uncertainties are very large, respectively 76--81\% and 83--90\%, owing essentially to poorly determined oscillator strengths and absorption cross sections. The average ratio of 1.8 obtained by using $\gamma_0$ is lowered to 1.3 when applying the inverted, squared errors as weights.

\begin{table*}
\caption{Upper limits for C$_{60}$:C$_{60}^+$ as $^{\mathrm{u}}N$(C$_{60}$)/$N$(C$_{60}^+$) from the comparison of the A$_1$ and $\gamma_0$ bands of C$_{60}$ with the 9577.5~{\AA} and 9632.7~{\AA} bands of C$_{60}^+$.}
\label{tbl:C60:C60cat}
\centering
\begin{tabular}{lllll}
\hline\hline
LOS & \multicolumn{2}{c}{C$_{60}$, A$_1$ (3980~{\AA})} & \multicolumn{2}{c}{C$_{60}$, $\gamma_0$ (6070~{\AA})} \\
 & C$_{60}^+$, 9577.5~{\AA} & C$_{60}^+$, 9632.7~{\AA} & C$_{60}^+$, 9577.5~{\AA} & C$_{60}^+$, 9632.7~{\AA} \\
\hline
HD~76341  & 6.2 (5.3)                   & 4.1 (3.6)                   & 3.1 (2.8)                     & 2.1 (1.9)                   \\
          & \textit{4.2} (\textit{3.3}) & \textit{2.8} (\textit{2.2}) & ...                           & ...                         \\
HD~136239 & 8.2 (6.5)                   & 9.1 (7.4)                   & 1.4 (1.2)                     & 1.6 (1.4)                   \\
          & \textit{3.3} (\textit{2.6}) & \textit{3.7} (\textit{2.9}) & \textit{1.4} (\textit{1.2})   & \textit{1.6} (\textit{1.4}) \\
HD~148379 & 7.3 (5.9)                   & 7.4 (6.0)                   & 1.2 (1.0)                     & 1.2 (1.0)                   \\
          & \textit{1.9} (\textit{1.5}) & \textit{1.9} (\textit{1.5}) & ...                           & ...                         \\
HD~167264 & 8.5 (7.5)                   & 4.1 (3.5)                   & 5.2 (4.7)                     & 2.5 (2.2)                   \\
          & \textit{4.3} (\textit{3.5}) & \textit{2.1} (\textit{1.7}) & ...                           & ...                         \\
HD~168625 & 2.1 (1.6)                   & 1.6 (1.3)                   & 1.3 (1.1)                     & 1.01 (0.88)                   \\
          & \textit{2.1} (\textit{1.6}) & \textit{1.6} (\textit{1.3}) & \textit{1.2} (\textit{1.0})   & \textit{0.90} (\textit{0.79}) \\
HD~169454 & 14 (12)                     & 8.6 (6.9)                   & 1.9 (1.7)                     & 1.1 (1.0)                   \\
          & \textit{6.9} (\textit{5.4}) & \textit{4.1} (\textit{3.2}) & \textit{1.6} (\textit{1.4})   & \textit{0.97} (\textit{0.85}) \\
HD~183143 & 4.0 (3.1)                   & 6.3 (5.1)                   & 1.3 (1.1)                     & 2.0 (1.8)                   \\
          & \textit{4.0} (\textit{3.1}) & \textit{6.3} (\textit{5.1}) & \textit{1.12} (\textit{0.97}) & \textit{1.8} (\textit{1.6}) \\
\hline
\end{tabular}
\tablefoot{Values between parentheses are computed errors. Values in italic were obtained by including in the continuum the fitted profiles of atomic stellar lines when in the 4000~{\AA} region, and the four DIBs nearest 6070~{\AA} and that of the nearby \ion{Ne}{I} line when in the 6000~{\AA} region.}
\end{table*}


\section{Discussion}\label{sec:discussion}

\subsection{The C$_{60}$:C$_{60}^+$ ratio in the diffuse ISM}\label{sec:dis-nt:cat}

The detection of buckminsterfullerene and its cation in the ISM has extended the network of physical and chemical mechanisms at work in that environment. The relative populations of gas-phase C$_{60}$, C$_{60}^+$, and C$_{60}^-$ would be primarily governed by the ISRF strength and the electron density, which are directly related to mechanisms such as photoionization of C$_{60}$ and electron detachment from C$_{60}^-$, electron attachment to C$_{60}$ and electron recombination with C$_{60}^+$.

\begin{figure}
\resizebox{8.8cm}{!}{\includegraphics{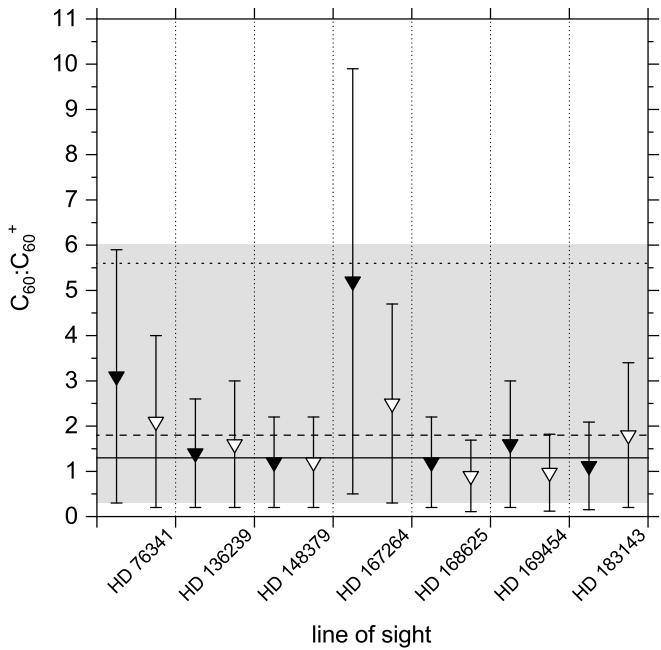}}
\caption{Values of C$_{60}$:C$_{60}^+$ in the diffuse ISM as a function of LOS (see Table~\ref{tbl:C60:C60cat}). Triangle symbols ($\blacktriangledown$ and $\triangledown$): upper limits $^{\mathrm{u}}N$(C$_{60}$, $\gamma_0$)/$N$(C$_{60}^+$, $\lambda$) with $\lambda$ = 9577.5~{\AA} and $\lambda$ = 9632.7~{\AA}, respectively. Error bars correspond to errors given in Table~\ref{tbl:C60:C60cat}. Gray area: C$_{60}$:C$_{60}^+$ according to \citet{Berne17}. Horizontal dotted line: C$_{60}$:C$_{60}^+$ for a typical diffuse medium as derived by \citet{Bakes95}. Horizontal dashed line: mean value of the present upper limits. Horizontal solid line: weighted mean value of the present upper limits. Vertical dotted lines separate the data relevant to each LOS.}
\label{fig:ratio}
\end{figure}

Figure~\ref{fig:ratio} illustrates $^{\mathrm{u}}N$(C$_{60}$, $\gamma_0$)/$N$(C$_{60}^+$) as a function of LOS and includes literature values of C$_{60}$:C$_{60}^+$ for comparison. The upper limits presently derived have an average value of 1.8, lowered to 1.3 when applying the inverted, squared errors as weights. Both averages are compatible with the interval of ratio values 0.3--6 presented by \citet{Berne17}, in particular its lower half. They suggest, however, that the ratio of 5.6 proposed by \citet{Bakes95} in their theoretical study is somewhat large.

The value proposed by \citeauthor{Bakes95} was determined for a typical diffuse cloud with 100~K gas temperature, 7.5~$\times$~10$^{-3}$~cm$^{-3}$ electron density, and 1~G$_0$ ISRF, conditions that may not be observed along the real LOSs presently studied. For instance, \citet{Gredel86} determined a gas temperature of 15~K, thus well below 100~K, from the observation of C$_2$ toward HD~169454. Additionally, in a study of C$_3$ observed toward the same star, \citet{Schmidt14} brought out two populations with excitation temperatures of 22.4 $\pm$ 1~K and 187 $\pm$ 25~K. Their analysis of the colder or less excited population led to an ISRF with a strength of 2 to 6~G$_0$, greater than the 1~G$_0$ ISRF used by \citeauthor{Bakes95}. As to electron density, \citet{Harrison13} proposed typical values in the range 0.01--0.06~cm$^{-3}$, up to eight times that adopted by \citeauthor{Bakes95}. Thus the conditions assumed by the latter authors may not be adequate to describe our LOSs, resulting through their interplay in a C$_{60}$:C$_{60}^+$ value greater than the weighted average upper limit derived in this study.

Moreover, properties of C$_{60}$ and its ions relevant to the determination of their relative populations are not accurately known. The C$_{60}^-$:C$_{60}$ and C$_{60}$:C$_{60}^+$ ratios are proportional, respectively, to the electron attachment rate of C$_{60}$ and the electron recombination rate of C$_{60}^+$. To date, both values are uncertain. \citet{Bakes95} attributed to neutral C$_{60}$ an electron attachment cross section of 10$^{-12}$ cm$^2$ in their typical diffuse medium, a value proposed by \citet{Lezius93}. Because the analysis of measurements for electrons with low energy is complex \citep{Kasperovich01,Lezius03}, the actual attachment cross section for an electron with an energy of $\sim$0.013~eV (corresponding to 100~K) may be different, and consequently C$_{60}^-$:C$_{60}$ too. As to the electron recombination rate of C$_{60}^+$, which has yet to be measured, \citeauthor{Bakes95} used a theoretical value extrapolated from a model for grains \citep{Draine87}. Yet they noted that this model gave rates an order of magnitude larger than measured ones when it was applied to the benzene (C$_6$H$_6$) and naphthalene (C$_{10}$H$_8$) cations. Thus the value they obtained for C$_{60}$:C$_{60}^+$ is possibly overestimated.

As to our result, one must be aware of the large uncertainties affecting the absorption cross sections and oscillator strengths we have employed to evaluate C$_{60}$:C$_{60}^+$. Any correction applied to one of them changes $^{\mathrm{u}}N$(C$_{60}$)/$N$(C$_{60}^+$), and C$_{60}$:C$_{60}^+$, in proportion.

The influence of the chosen oscillator strengths on the C$_{60}$:C$_{60}^+$ ratio is illustrated by the following evaluations, which differs from ours. \citet{Fulara93} had estimated for the two bands of C$_{60}^+$ a total oscillator strength in the range 0.003--0.006, that is, five to nine times smaller than the value of 0.0278 we have adopted (from adding 0.0164 to 0.0114, see Sect.~\ref{sec:laboscat}). The former value has been used by \citet{Herbig00} to estimate $N$(C$_{60}^+$) toward Cyg OB2/8A, which, combined with an evaluation of $^{\mathrm{u}}N$(C$_{60}$) lower than ours by an order of magnitude, led to C$_{60}$:C$_{60}^+$ being 0.0075 at most. In that study, the author mentioned his doubts concerning the result and explained how the value for $^{\mathrm{u}}N$(C$_{60}$) might be incorrect owing to a possibly much underestimated FWHM (see Sect.~\ref{sec:dis-searches}). In a second case, \citeauthor{Maier17} \citep{Maier17,Campbell17} inevitably derived a high degree of ionization, 98\%, because they borrowed $^{\mathrm{u}}N$(C$_{60}$) from \citeauthor{Herbig00}. They also stated that the degree of ionization should be at least 90\% even if one of the DIBs observed near 6000~{\AA} was absorption by C$_{60}$ \citep{Maier17}. The value was derived from elements in a discussion by \citet{Omont16}, among which a theoretical oscillator strength of 0.01 attributed to the potential C$_{60}$ DIB, a value 25 times greater than the strength presently determined for the $\gamma_0$ band. We remark that abundances of neutral molecules and cations as evaluated by \citeauthor{Omont16} would give 80\% as the degree of ionization.

Finally, when using Eqs.~\ref{equ:N-Gauss} and \ref{equ:uN}, we assume that the area of the band is a valid measurement of the population of the absorbing species in absence of vibrational excitation, and that it is conserved over the range of temperature observed in diffuse and translucent clouds. Taking into account the conditions of the laboratory measurements on C$_{60}$ and C$_{60}^+$, the bands that interest us represent transitions from the electronic ground state with zero quantum of vibrational excitation. As a transfer of population to excited vibrational levels occurs when the temperature increases, the validity of the computed column densities may be affected when looking at the warmer regions of diffuse clouds. Nevertheless, provided the vibrational modes of the initial and final states of a transition are similar, the loss of area caused by an increase of the temperature may be compensated by the rise of overlapping bands that correspond to transitions from the thermally populated vibrational levels with conservation of the vibrational state. In this case the C$_{60}$:C$_{60}^+$ ratio may be valid for a range of temperatures that should be determined.

We have mentioned the observation by \citet{IglesiasGroth19} of interstellar fullerenes C$_{60}$ and C$_{70}$, and the ions C$_{60}^+$ and C$_{60}^-$, in IR emission spectra of the star-forming region IC~348. Ion fractions of 20\% and 10\% were determined for C$_{60}^+$ and C$_{60}^-$, respectively, giving a C$_{60}$:C$_{60}^+$ ratio of 3.5. We do not attempt a comparison with our results because the ISRF in the targeted regions is probably much stronger than in the typical diffuse or translucent interstellar clouds we are interested in. Actually \citeauthor{IglesiasGroth19} adopted ISRF strengths of 20 and 45~G$_0$ in their analysis. Moreover we are not aware of an evaluation of the electron density in the relevant regions. This value would have to be taken into account \citep{Bakes94,Bakes95}.

\subsection{Searches for vibronic bands of C$_{60}$ in the diffuse ISM}\label{sec:dis-searches}

Searches for interstellar C$_{60}$ were attempted following the report of a narrow absorption by C$_{60}$ at 3860~{\AA} \citep{Snow89,Somerville89,Somerville93}. The measurements concerned actually van der Waals complexes of C$_{60}$ with benzene and dichloromethane, from which the band of free C$_{60}$ was predicted and attributed to the origin of the 1$^1$T$_{1u} \leftarrow ^1$A$_g$ transition \citep{Heath87}. Because it is firmly established that the latter is found at 4024~{\AA} as the A$_0$ band, the upper limit, on the order of 10$^{14}$~cm$^{-2}$, derived by \citet{Snow89} for $N$(C$_{60}$) required a revision that \citet{Herbig00} attempted without obtaining a satisfying result.

\citet{Ehrenfreund97} searched for absorption by C$_{60}$ near 3980~{\AA} and near 6070~{\AA} in a spectrum measured toward BD+63$^{\circ}$1964. They found a weak band with 20~m{\AA} equivalent width close to 3980~{\AA}. At 6070~{\AA}, with a detection limit of 3~m{\AA} equivalent width, the spectrum did not show any absorption band. A possible DIB observed at 6220~{\AA} appeared to coincide with a band seen beside the $\gamma_0$ band in the spectrum of C$_{60}$ \citep{Haufler91}. Because laboratory spectra indicate that the $\gamma_0$ band is the strongest component in the pair, the DIB at 6220~{\AA} is likely caused by an interstellar substance that is not C$_{60}$. Additionally, the wavelengths of the bands observed by \citeauthor{Ehrenfreund97} did not coincide exactly with those derived from laboratory spectra of C$_{60}$ \citep{Herbig00}.

\citet{Herbig00} looked for various bands of C$_{60}$ toward several objects. Using band C reported by \citet{Leach92} at 3284~{\AA} in hexane at 300~K, an upper limit of 4.5~$\times$~10$^{11}$~cm$^{-2}$ was found for $N$(C$_{60}$) toward Cyb~OB2/8A. The value is invalid, as actually suspected by the author, because it was derived by hypothesizing an FWHM of 1~{\AA} for band C in the gas phase, which is too low a value considering that the band has a width of 1660~cm$^{-1}$ in the spectrum of C$_{60}$ in Ne ice \citep{Sassara01b}, that is, 173~{\AA} (see also Fig.~\ref{fig:MIS}). Using this underestimated FWHM to evaluate $^{\mathrm{u}}N$(C$_{60}$, C), \citeauthor{Herbig00} obtained a C$_{60}$:C$_{60}^+$ ratio on the order of 0.01 at maximum (see Sect.~\ref{sec:dis-nt:cat}).

\citet{GarciaHernandez12}, \citet{GarciaHernandez13a}, and \citet{DiazLuis15} sought C$_{60}$ toward hot star DY~Cen, and planetary nebulae Tc~1 and IC~418. From 1$\sigma$ equivalent-width detection limits at 3980~{\AA}, the authors derived upper limits of 1.5~$\times$~10$^{13}$, 1~$\times$~10$^{13}$, and 4~$\times$~10$^{13}$~cm$^{-2}$ for the column densities of C$_{60}$ toward DY~Cen, Tc~1, and IC~418, respectively. The $^{\mathrm{u}}N$(C$_{60}$, A$_1$) values we have obtained are of the same order (Table~\ref{tbl:C60}), though we took into account three equivalent-width detection thresholds in the form of $^{\mathrm{u}}W$. Analysis using the $\gamma_0$ band has given lower values, of the same order, however, yet by taking into account three equivalent-width detection thresholds again.

Past searches for interstellar C$_{60}$ focused on the detection of its UV and near-UV bands. To our knowledge, only \citet{Ehrenfreund97} and \citet{Herbig00} examined the possibility that DIBs near 6070~{\AA} could be related to C$_{60}$. Because it is narrow and lies at a visible wavelength, the $\gamma_0$ band is a more convenient target than, for instance, A$_1$, as demonstrated by the $^{\mathrm{u}}W$ values in Table~\ref{tbl:C60}, despite the larger uncertainties that affect the description of this absorption. Furthermore, spectra in rare-gas matrices have shown that the $\gamma_2$ band is similar both in strength and width to $\gamma_0$ (\citealt{Gasyna91}, \citealt{Sassara97}, and this work). Solvatochromism measurements \citep{Catalan94,Renge95} and low-temperature gas-phase spectroscopy \citep{Hansen97} place $\gamma_2$ at $\sim$584~nm. It is desirable that a future search for C$_{60}$ targets both bands. Already surveys list DIBs at positions close to those of the $\gamma_0$ and $\gamma_2$ bands. A comparison is not straightforward, however, because of the uncertainties affecting the laboratory data \citep[see also][]{Herbig00}. Our simple analysis of DIB measurements toward three targets \citep{Tuairisg00} indicates that the most strongly correlated DIBs near the positions of $\gamma_0$ and $\gamma_2$ are the bands at 5828.56~{\AA} and 6068.45~{\AA}. Not taking uncertainties into account, the separation is 10~{\AA} too large for assigning the pair to the $\gamma_0$ and $\gamma_2$ bands.


\section{Conclusions}
After reevaluating the oscillator strengths of bands in the absorption spectrum of C$_{60}$, we exploited spectra of background stars and available DIB measurements to estimate the C$_{60}$:C$_{60}^+$ ratio in diffuse and translucent clouds. We have obtained an average upper limit of $\sim$1.3 corresponding to a minimum degree of ionization of $\sim$40\%. Combining observed emission and absorption bands of fullerene C$_{60}$ and its cation, respectively, \citet{Berne17} determined a C$_{60}$:C$_{60}^+$ ratio equal to 0.3--6 in the diffuse ISM. The two results present common values and they suggest that the fraction of neutral molecules in the fullerene population of the diffuse ISM may be notable even though its electronic transitions have not been detected. Our approach allows us, however, not to rely on IR emission spectra, which present a degree of ambiguity as to the species that cause them. As to the difference between our result and the theoretical prediction by \citet{Bakes95}, it can simply be the consequence of inadequate values given to parameters of the model. In that respect, more accurate measurements of interstellar absorption at near-UV and visible wavelengths would be useful to constrain the local ISRF strength and electron density. Furthermore, the comparison of these measurements with those of mid-IR emission bands would allow us to distinguish the part of emission in diffuse and translucent clouds that is actually caused by gas-phase C$_{60}$ molecules.

The very large uncertainties of the upper limits we have derived are caused by the lack of accurate experimental data. New experiments are needed in order to better determine the absorption cross sections, oscillator strengths, and FWHMs of the various bands involved in this study, in particular that of the $\gamma_0$ band of C$_{60}$ at 6070~{\AA}. Because it is narrow and arises in a wavelength domain favorable to observations, the $\gamma_0$ band may be used to obtain the lowest equivalent-width detection threshold values. A better characterization of the $\gamma_2$ band is also desirable. Since matrix-isolation measurements have shown its strength and width are similar to those of $\gamma_0$, a search for interstellar C$_{60}$ at visible wavelengths would be considered successful if it revealed both bands. Another way to confirm the detection of $\gamma_0$ would be to observe the near-UV A$_1$ band too, which is especially worth attempting when the spectral type of the target star is favorable.

Finally we need electronic spectra of C$_{60}^-$ at low temperature in the gas phase in order to progress. To date, measurements of vibronic bands of the anion are scarce and their resolution modest \citep{Tomita05,Stochkel13}.


\begin{acknowledgements}
The authors gratefully acknowledge the use of observations collected at the European Southern Observatory under ESO programs 079.C-0597 (R. Gredel, for HD~169454 and HD~183143) 081.C-0475 (J. Smoker, for HD~76341, HD~136239, and HD~167264), 082.C-0566 (Y. Beletsky, for HD~136239 and HD~148379), 194.C-0833 (N. Cox, for HD~167264, HD~169454, and HD~183143), and 266.D-5655 (Paranal Observatory, ESO, for HD~168625). S.K. is thankful to the Deutsche Forschungsgemeinschaft (DFG) for its support through project No. 413610339. Special thanks are owed to R. Gredel for discussions on the wavelength calibration of astronomical spectra and for suggesting analysis methods, and to D. Strelnikov for exchanges on the determination of oscillator strengths.
\end{acknowledgements}


\bibliographystyle{aa}
\bibliography{roui2108}

\begin{thebibliography}{78}
\expandafter\ifx\csname natexlab\endcsname\relax\def\natexlab#1{#1}\fi

\bibitem[{Bakes \& Tielens(1994)}]{Bakes94}
Bakes, E. L.~O. \& Tielens, A. G. G.~M. 1994, {ApJ}, 427, 822

\bibitem[{Bakes \& Tielens(1995)}]{Bakes95}
Bakes, E. L.~O. \& Tielens, A. G. G.~M. 1995, in Astrophysics and Space Science
  Library, Vol. 202, {T}he {D}iffuse {I}nterstellar {B}ands, ed. A.~G. G.~M.
  Tielens \& T.~P. Snow (Dordrecht: Springer), 315--321

\bibitem[{Bern\'e {et~al.}(2017)Bern\'e, Cox, Mulas, \& Joblin}]{Berne17}
Bern\'e, O., Cox, N. L.~J., Mulas, G., \& Joblin, C. 2017, {A}\&{A}, 605, L1

\bibitem[{Brieva {et~al.}(2016)Brieva, Gredel, J{\"a}ger, Huisken, \&
  Henning}]{Brieva16}
Brieva, A.~C., Gredel, R., J{\"a}ger, C., Huisken, F., \& Henning, T. 2016,
  {ApJ}, 826, 122

\bibitem[{Cami(2014)}]{Cami14}
Cami, J. 2014, in Proceedings of the International Astronomical Union, Vol.~9,
  {T}he {D}iffuse {I}nterstellar {B}ands, ed. J.~Cami \& N.~Cox (Cambridge:
  Cambridge University Press), 370--374

\bibitem[{Cami {et~al.}(2010)Cami, Bernard-Salas, Peeters, \& Malek}]{Cami10}
Cami, J., Bernard-Salas, J., Peeters, E., \& Malek, S.~E. 2010, {Sci}, 329,
  1180

\bibitem[{Campbell {et~al.}(2015)Campbell, Holz, Gerlich, \&
  Maier}]{Campbell15}
Campbell, E.~K., Holz, M., Gerlich, D., \& Maier, J.~P. 2015, {Natur}, 523, 322

\bibitem[{Campbell {et~al.}(2016a)Campbell, Holz, \& Maier}]{Campbell16a}
Campbell, E.~K., Holz, M., \& Maier, J.~P. 2016a, {ApJL}, 826, L4

\bibitem[{Campbell {et~al.}(2016b)Campbell, Holz, Maier, Gerlich, Walker, \&
  Bohlender}]{Campbell16b}
Campbell, E.~K., Holz, M., Maier, J.~P., {et~al.} 2016b, {ApJ}, 822, 17

\bibitem[{Campbell \& Maier(2017)}]{Campbell17}
Campbell, E.~K. \& Maier, J.~P. 2017, {JChPh}, 146, 160901

\bibitem[{Cardelli {et~al.}(1989)Cardelli, Clayton, \& Mathis}]{Cardelli89}
Cardelli, J.~A., Clayton, G.~C., \& Mathis, J.~S. 1989, {ApJ}, 345, 245

\bibitem[{Carleer {et~al.}(1999)Carleer, Jenouvrier, Vandaele, Bernath,
  M{\'e}rienne, Colin, Zobov, Polyansky, Tennyson, \& Savin}]{Carleer99}
Carleer, M., Jenouvrier, A., Vandaele, A.-C., {et~al.} 1999, {JChPh}, 111, 2444

\bibitem[{Cat{\'a}lan(1994)}]{Catalan94}
Cat{\'a}lan, J. 1994, {CPL}, 223, 159

\bibitem[{Cat{\'a}lan \& P{\'e}rez(2002)}]{Catalan02}
Cat{\'a}lan, J. \& P{\'e}rez, P. 2002, {FNCN}, 10, 171

\bibitem[{Chupka \& Klots(1997)}]{Chupka97}
Chupka, W.~A. \& Klots, C.~E. 1997, {IJMSI}, 167/168, 595

\bibitem[{Close {et~al.}(1997)Close, Federmann, Hoffmann, \& Quaas}]{Close97}
Close, J.~D., Federmann, F., Hoffmann, K., \& Quaas, N. 1997, {CPL}, 276, 393

\bibitem[{Cordiner {et~al.}(2019)Cordiner, Linnartz, , Cox, Cami, , Najarro,
  Proffitt, Lallement, Ehrenfreund, Foing, Gull, Sarre, \&
  Charnley}]{Cordiner19}
Cordiner, M.~A., Linnartz, H., , {et~al.} 2019, {ApJL}, 875, L28

\bibitem[{D{\'i}az-Luis {et~al.}(2015)D{\'i}az-Luis, Garc{\'i}a-Hern{\'a}ndez,
  Rao, Manchado, \& Cataldo}]{DiazLuis15}
D{\'i}az-Luis, J.~J., Garc{\'i}a-Hern{\'a}ndez, D.~A., Rao, N.~K., Manchado,
  A., \& Cataldo, F. 2015, {A}\&{A}, 573, A97

\bibitem[{Draine \& Sutin(1987)}]{Draine87}
Draine, B.~T. \& Sutin, B. 1987, {ApJ}, 320, 803

\bibitem[{Edwards \& Leach(1993)}]{Edwards93}
Edwards, S.~A. \& Leach, S. 1993, {A}\&{A}, 272, 533

\bibitem[{Ehrenfreund \& Foing(1997)}]{Ehrenfreund97}
Ehrenfreund, P. \& Foing, B.~H. 1997, {AdSpR}, 19, 1033

\bibitem[{Fan {et~al.}(2019)Fan, Hobbs, Dahlstrom, Welty, York, Rachford, Snow,
  Sonnentrucker, Baskes, \& Zhao}]{Fan19}
Fan, H., Hobbs, L.~M., Dahlstrom, J.~A., {et~al.} 2019, {ApJ}, 878, 151

\bibitem[{Foing \& Ehrenfreund(1994)}]{Foing94}
Foing, B.~H. \& Ehrenfreund, P. 1994, {Natur}, 369, 296

\bibitem[{Foing \& Ehrenfreund(1997)}]{Foing97}
Foing, B.~H. \& Ehrenfreund, P. 1997, {A}\&{A}, 319, L59

\bibitem[{Fulara {et~al.}(1993)Fulara, Jakobi, \& Maier}]{Fulara93}
Fulara, J., Jakobi, M., \& Maier, J.~P. 1993, {CPL}, 211, 227

\bibitem[{Galazutdinov {et~al.}(2015)Galazutdinov, Kre{\l}owski, Beletsky, \&
  Valyavin}]{Galazutdinov15}
Galazutdinov, G., Kre{\l}owski, J., Beletsky, Y., \& Valyavin, G. 2015, {PASP},
  127, 356

\bibitem[{Galazutdinov {et~al.}(2000)Galazutdinov, Musaev, Kre{\l}owski, \&
  Walker}]{Galazutdinov00b}
Galazutdinov, G.~A., Musaev, F.~A., Kre{\l}owski, J., \& Walker, G. A.~H. 2000,
  {PASP}, 112, 648

\bibitem[{Galazutdinov {et~al.}(2017)Galazutdinov, Shimansky, Bondar, Valyavin,
  \& Kre{\l}owski}]{Galazutdinov17}
Galazutdinov, G.~A., Shimansky, V.~V., Bondar, A., Valyavin, G., \&
  Kre{\l}owski, J. 2017, {MNRAS}, 465, 3956

\bibitem[{Garc{\'i}a-Hern{\'a}ndez {et~al.}(2013)Garc{\'i}a-Hern{\'a}ndez,
  Cataldo, \& Manchado}]{GarciaHernandez13b}
Garc{\'i}a-Hern{\'a}ndez, D.~A., Cataldo, F., \& Manchado, A. 2013, {MNRAS},
  434, 415

\bibitem[{Garc{\'i}a-Hern{\'a}ndez \& D{\'i}az-Luis(2013)}]{GarciaHernandez13a}
Garc{\'i}a-Hern{\'a}ndez, D.~A. \& D{\'i}az-Luis, J.~J. 2013, {A}\&{A}, 550, L6

\bibitem[{Garc{\'i}a-Hern{\'a}ndez {et~al.}(2012)Garc{\'i}a-Hern{\'a}ndez, Rao,
  \& Lambert}]{GarciaHernandez12}
Garc{\'i}a-Hern{\'a}ndez, D.~A., Rao, N.~K., \& Lambert, D.~L. 2012, {ApJL},
  759, L21

\bibitem[{Gasyna {et~al.}(1991)Gasyna, Schatz, Hare, Dennis, Kroto, Taylor, \&
  Walton}]{Gasyna91}
Gasyna, Z., Schatz, P.~N., Hare, J.~P., {et~al.} 1991, {CPL}, 183, 283

\bibitem[{Gredel \& M\"unch(1986)}]{Gredel86}
Gredel, R. \& M\"unch, G. 1986, {A}\&{A}, 154, 336

\bibitem[{Hansen {et~al.}(1997)Hansen, M{\"u}ller, Brockhaus, Campbell, \&
  Hertel}]{Hansen97}
Hansen, K., M{\"u}ller, R., Brockhaus, P., Campbell, E. E.~B., \& Hertel, I.~V.
  1997, {ZPhyD}, 42, 153

\bibitem[{Harrison {et~al.}(2013)Harrison, Faure, \& Tennyson}]{Harrison13}
Harrison, S., Faure, A., \& Tennyson, J. 2013, {MNRAS}, 435, 3541

\bibitem[{Haufler {et~al.}(1991)Haufler, Chai, Chibante, Fraelich, Weisman,
  Curl, \& Smalley}]{Haufler91}
Haufler, R.~E., Chai, Y., Chibante, L. P.~F., {et~al.} 1991, {JChPh}, 95, 2197

\bibitem[{Heath {et~al.}(1987)Heath, Curl, \& Smalley}]{Heath87}
Heath, J.~R., Curl, R.~F., \& Smalley, R.~E. 1987, {JChPh}, 87, 4236

\bibitem[{Herbig(2000)}]{Herbig00}
Herbig, G.~H. 2000, {ApJ}, 542, 334

\bibitem[{Hobbs {et~al.}(2009)Hobbs, York, Thorburn, Snow, Bishof, Friedman,
  McCall, Oka, Rachford, \& Sonnentrucker}]{Hobbs09}
Hobbs, L.~M., York, D.~G., Thorburn, J.~A., {et~al.} 2009, {ApJ}, 705, 32

\bibitem[{Hora {et~al.}(1996)Hora, P{\'a}nek, Navr{\'a}til,
  Handl{\'i}{\v{r}}ov{\'a}, Huml{\'i}{\v{c}}ek, Sitter, \& Stifter}]{Hora96}
Hora, J., P{\'a}nek, P., Navr{\'a}til, K., {et~al.} 1996, {PhRvB}, 54, 1478

\bibitem[{Hunter {et~al.}(2006)Hunter, Smoker, Keenan, Ledoux, Jehin, Cabanac,
  Melo, \& Bagnulo}]{Hunter06}
Hunter, I., Smoker, J.~V., Keenan, F.~P., {et~al.} 2006, {MNRAS}, 367, 1478

\bibitem[{Iglesias-Groth(2019)}]{IglesiasGroth19}
Iglesias-Groth, S. 2019, {MNRAS}, 489, 1509

\bibitem[{Jannuzi {et~al.}(1988)Jannuzi, Black, Lada, \& {van
  Dishoeck}}]{Jannuzi88}
Jannuzi, B.~T., Black, J.~H., Lada, C.~J., \& {van Dishoeck}, E.~F. 1988,
  {ApJ}, 332, 995

\bibitem[{Kasperovich {et~al.}(2001)Kasperovich, Tikhonov, \&
  Kresin}]{Kasperovich01}
Kasperovich, V., Tikhonov, G., \& Kresin, V.~V. 2001, {CPL}, 337, 55

\bibitem[{Krasnokutski {et~al.}(2019)Krasnokutski, Gruenewald, J{\"a}ger, Otto,
  Forker, Fritz, \& Henning}]{Krasnokutski19}
Krasnokutski, S.~A., Gruenewald, M., J{\"a}ger, C., {et~al.} 2019, {ApJ}, 874,
  149

\bibitem[{Lawton {et~al.}(2008)Lawton, Churchill, York, Ellison, Snow, Johnson,
  Ryan, \& Benn}]{Lawton08}
Lawton, B., Churchill, C.~W., York, B.~A., {et~al.} 2008, {AJ}, 136, 994

\bibitem[{Leach {et~al.}(1992)Leach, Vervloet, Despr{\`e}s, Br{\'e}heret, Hare,
  Dennis, Kroto, Taylor, \& Walton}]{Leach92}
Leach, S., Vervloet, M., Despr{\`e}s, A., {et~al.} 1992, {CP}, 160, 451

\bibitem[{L{\'e}ger {et~al.}(1988)L{\'e}ger, {d'Hendecourt}, Verstraete, \&
  Schmidt}]{Leger88}
L{\'e}ger, A., {d'Hendecourt}, L., Verstraete, L., \& Schmidt, W. 1988,
  {A}\&{A}, 203, 145

\bibitem[{Lezius(2003)}]{Lezius03}
Lezius, M. 2003, {IJMSp}, 223-224, 447

\bibitem[{Lezius {et~al.}(1993)Lezius, Scheier, \& M{\"a}rk}]{Lezius93}
Lezius, M., Scheier, P., \& M{\"a}rk, T.~D. 1993, {CPL}, 203, 232

\bibitem[{Maier \& Campbell(2017)}]{Maier17}
Maier, J.~P. \& Campbell, E.~K. 2017, Angew. Chem. Int. Ed., 56, 4920

\bibitem[{Mulliken(1939)}]{Mulliken39}
Mulliken, R.~S. 1939, {JChPh}, 7, 14

\bibitem[{Negri {et~al.}(1992)Negri, Orlandi, \& Zerbetto}]{Negri92}
Negri, F., Orlandi, G., \& Zerbetto, F. 1992, {JChPh}, 97, 6496

\bibitem[{Omont(2016)}]{Omont16}
Omont, A. 2016, {A}\&{A}, 590, A52

\bibitem[{Orlandi \& Negri(2002)}]{Orlandi02}
Orlandi, G. \& Negri, F. 2002, Photochem. Photobiol. Sci., 1, 289

\bibitem[{Pan {et~al.}(2005)Pan, Federman, Sheffer, \& Andersson}]{Pan05}
Pan, K., Federman, S.~R., Sheffer, Y., \& Andersson, B.-G. 2005, {ApJ}, 633,
  986

\bibitem[{Renge(1995)}]{Renge95}
Renge, I. 1995, {JPhCh}, 99, 15955

\bibitem[{Rice {et~al.}(2018)Rice, Federman, Flagey, Goldsmith, Langer, Pineda,
  \& Lambert}]{Rice18}
Rice, J.~S., Federman, S.~R., Flagey, N., {et~al.} 2018, {ApJ}, 858, 111

\bibitem[{Rouill{\'e} {et~al.}(2009)Rouill{\'e}, Steglich, Huisken, Henning, \&
  M{\"u}llen}]{Rouille09}
Rouill{\'e}, G., Steglich, M., Huisken, F., Henning, T., \& M{\"u}llen, K.
  2009, {JChPh}, 131, 204311

\bibitem[{Sassara {et~al.}(2001)Sassara, Zerza, Chergui, \& Leach}]{Sassara01b}
Sassara, A., Zerza, G., Chergui, M., \& Leach, S. 2001, {ApJS}, 135, 263

\bibitem[{Sassara {et~al.}(1997)Sassara, Zerza, Chergui, Negri, \&
  Orlandi}]{Sassara97}
Sassara, A., Zerza, G., Chergui, M., Negri, F., \& Orlandi, G. 1997, {JChPh},
  107, 8731

\bibitem[{Schmidt {et~al.}(2014)Schmidt, Kre{\l}owski, Galazutdinov, Zhao,
  Haddad, Ubachs, \& Linnartz}]{Schmidt14}
Schmidt, M.~R., Kre{\l}owski, J., Galazutdinov, G.~A., {et~al.} 2014, {MNRAS},
  441, 1134

\bibitem[{Seiler(1987)}]{Seiler87}
Seiler, F.~A. 1987, {R}isk {A}nal., 7, 509

\bibitem[{Sellgren {et~al.}(2010)Sellgren, Werner, Ingalls, Smith, Carleton, \&
  Joblin}]{Sellgren10}
Sellgren, K., Werner, M.~W., Ingalls, J.~G., {et~al.} 2010, {ApJL}, 722, L54

\bibitem[{Snow \& {McCall}(2006)}]{Snow06}
Snow, T.~P. \& {McCall}, B.~J. 2006, {ARA}\&{A}, 44, 367

\bibitem[{Snow \& Seab(1989)}]{Snow89}
Snow, T.~P. \& Seab, C.~G. 1989, {A}\&{A}, 213, 291

\bibitem[{Somerville \& Bellis(1989)}]{Somerville89}
Somerville, W.~B. \& Bellis, J.~G. 1989, {MNRAS}, 240, 41

\bibitem[{Somerville \& Crawford(1993)}]{Somerville93}
Somerville, W.~B. \& Crawford, I.~A. 1993, {FaTr}, 89, 2261

\bibitem[{Sowers {et~al.}(1972)Sowers, Williams, Hamm, \& Arakawa}]{Sowers72}
Sowers, B.~L., Williams, M.~W., Hamm, R.~N., \& Arakawa, E.~T. 1972, {JChPh},
  57, 167

\bibitem[{St{\o}chkel \& Andersen(2013)}]{Stochkel13}
St{\o}chkel, K. \& Andersen, J.~U. 2013, {JChPh}, 139, 164304

\bibitem[{Strelnikov {et~al.}(2015)Strelnikov, Kern, \& Kappes}]{Strelnikov15}
Strelnikov, D., Kern, B., \& Kappes, M.~M. 2015, {A}\&{A}, 584, A55

\bibitem[{Thompson {et~al.}(2008)Thompson, Keenan, Dufton, Trundle, Ryans, \&
  Crowther}]{Thompson08}
Thompson, H. M.~A., Keenan, F.~P., Dufton, P.~L., {et~al.} 2008, {MNRAS}, 383,
  729

\bibitem[{Tomita {et~al.}(2005)Tomita, Andersen, Bonderup, Hvelplund, Liu,
  Nielsen, Pedersen, Rangama, Hansen, \& Echt}]{Tomita05}
Tomita, S., Andersen, J.~U., Bonderup, E., {et~al.} 2005, {PhRvL}, 94, 053002

\bibitem[{Tuairisg {et~al.}(2000)Tuairisg, Cami, Foing, Sonnentrucker, \&
  Ehrenfreund}]{Tuairisg00}
Tuairisg, S.~{\'O}., Cami, J., Foing, B.~H., Sonnentrucker, P., \& Ehrenfreund,
  P. 2000, {A}\&{AS}, 142, 225

\bibitem[{Walker {et~al.}(2015)Walker, Bohlender, Maier, \&
  Campbell}]{Walker15}
Walker, G. A.~H., Bohlender, D.~A., Maier, J.~P., \& Campbell, E.~K. 2015,
  {ApJL}, 812, L8

\bibitem[{Walker {et~al.}(2017)Walker, Campbell, Maier, \&
  Bohlender}]{Walker17}
Walker, G. A.~H., Campbell, E.~K., Maier, J.~P., \& Bohlender, D. 2017, {ApJ},
  843, 56

\bibitem[{Walker {et~al.}(2016)Walker, Campbell, Maier, Bohlender, \&
  Malo}]{Walker16}
Walker, G. A.~H., Campbell, E.~K., Maier, J.~P., Bohlender, D., \& Malo, L.
  2016, {ApJ}, 831, 130

\bibitem[{Weselak {et~al.}(2000)Weselak, Schmidt, \& Kre{\l}owski}]{Weselak00}
Weselak, T., Schmidt, M., \& Kre{\l}owski, J. 2000, {A}\&{AS}, 142, 239

\end{thebibliography}


\begin{appendix}


\section{Observational data}\label{apx:data}

For each LOS and wavelength domain examined in the present study, Table~\ref{tbl:obsdata} gives the resolution of the observational spectra, the angular distance to the target star, and the data sets. Information on the LOSs such as the spectral type of the target star and the observed color excess can be found for instance in \citet{Hunter06} for HD~76341, HD~136239, HD~148379, HD~167264, and HD~169454, and in \citet{Fan19} for HD~168625 and HD~183143. Selected information on each target star is also found in \citet{Galazutdinov17}.

\begin{table*}
\caption{Observational data sets.}
\label{tbl:obsdata}
\centering
\begin{tabular}{lllll}
\hline\hline
LOS & Region ({\AA}) & Resolution & Distance (") & Summed Data Sets \\
\hline
HD~76341  & 4000 & 65030 & 0.316 & ADP.2013-09-27T12$\_$26$\_$24.837.fits, *$\_$25.510.fits, *$\_$25.653.fits, \\
          &      &       &       & *$\_$26.350.fits, *$\_$26.810.fits \\
          & 6000 & 74450 & 0.316 & ADP.2013-09-27T12$\_$26$\_$24.790.fits, *$\_$24.947.fits, *$\_$24.960.fits, \\
					&      &       &       & *$\_$25.710.fits, *$\_$26.303.fits \\
HD~136239 & 4000 & 77810 & 0.423 & ADP.2013-09-27T17$\_$01$\_$53.550.fits, *$\_$53.990.fits \\
          & 6000 & 74450 & 0.305 & ADP.2013-09-27T12$\_$26$\_$24.710.fits, *$\_$24.900.fits, *$\_$25.030.fits, \\
					&      &       &       & *$\_$25.090.fits, *$\_$25.723.fits, *$\_$26.530.fits, *$\_$26.593.fits \\
HD~148379 & 4000 & 77810 & 0.639 & ADP.2013-09-27T17$\_$10$\_$35.883.fits, *$\_$35.897.fits, *$\_$36.043.fits, \\
          &      &       &       & *$\_$36.157.fits, *$\_$36.217.fits, *$\_$36.503.fits, *$\_$36.650.fits, \\
					&      &       &       & *$\_$36.777.fits, *$\_$36.930.fits, *$\_$37.050.fits \\
          & 6000 & 87410 & 0.613 & ADP.2013-09-27T17$\_$10$\_$35.643.fits, *$\_$35.723.fits, *$\_$35.750.fits, \\
					&      &       &       & *$\_$35.890.fits, *$\_$36.297.fits, *$\_$36.343.fits, *$\_$36.517.fits, \\
					&      &       &       & *$\_$36.750.fits, *$\_$36.797.fits, *$\_$37.257.fits \\
HD~167264 & 4000 & 71050 & 1.23  & ADP.2016-06-13T12$\_$13$\_$59.824.fits, *.825.fits, *.826.fits, *.828.fits, \\
					&      &       &       & *.830.fits, *.832.fits, *.834.fits, *.835.fits, *.837.fits, *.842.fits, \\
					&      &       &       & *.847.fits, *.851.fits \\
          & 6000 & 74450 & 0.383 & ADP.2013-09-27T13$\_$08$\_$45.867.fits, *$\_$45.880.fits, *$\_$45.920.fits, \\
					&      &       &       & *$\_$45.993.fits, *$\_$46.000.fits, *$\_$46.007.fits, \\
					&      &       &       & *$\_$46.013.fits, *$\_$46.113.fits, *$\_$46.160.fits \\
HD~168625 & 4000 & 65030 & 4.46  & ADP.2017-10-24T09$\_$50$\_$31.392.fits, *.394.fits, *.396.fits, *.412.fits \\
          & 6000 & 74450 & 0.027 & ADP.2017-10-24T09$\_$50$\_$31.415.fits, *.437.fits \\
HD~169454 & 4000 & 71050 & 0.511 & ADP.2016-09-06T06$\_$22$\_$18.210.fits, *.232.fits, *.234.fits \\
          & 6000 & 66320 & 0.451 & ADP.2017-10-20T16$\_$04$\_$39.935.fits, *.946.fits, *.947.fits, *.948.fits, \\
					&      &       &       & *.951.fits, *.952.fits, *.955.fits, *.969.fits, *.981.fits, *.982.fits \\
HD~183143 & 4000 & 71050 & 0.827 & ADP.2018-10-11T06$\_$09$\_$48.104.fits \\
          & 6000 & 66320 & 0.31  & ADP.2013-09-26T22$\_$20$\_$55.787.fits, *.807.fits, *.820.fits \\
\hline
\end{tabular}
\tablefoot{Asterisks replace common parts in successive file names. The ESO has reprocessed the UVES datasets and stacked them when adequate during the preparation of this manuscript. The original datasets are still valid.}
\end{table*}


\section{Wavelengths and standards of rest}\label{apx:shifts}

In this work, wavelengths are expressed in units of angstrom ({\AA}) when explicit information shows that they were measured in air. Wavelengths are given in units of nanometer (nm) whenever the calibration conditions are not indicated in the source material. Wavenumbers in reciprocal centimeter (cm$^{-1}$) may be for air or vacuum.

For each LOS we have examined lines of identified interstellar species to evaluate their radial velocity relative to the observer, that is, the wavelength standard of rest in the archival spectra. This information was used to change this standard for an interstellar absorber plausibly coexisting with fullerenes, allowing us to locate bands of interstellar C$_{60}$ in a straightforward manner by using the laboratory rest wavelengths.

We are looking for interstellar C$_{60}$ in diffuse and translucent molecular clouds, for which \ion{K}{1}, CH, and CH$^+$ are the preferred tracers of the less dense regions, and CN of the denser ones \citep{Pan05,Rice18}. In some cases, DIBs exhibit a shift relative to the lines of these tracers, indicating that cloud structures may be complex \citep{Galazutdinov15}. Red and blue shifts as great as $\sim$10~km s$^{-1}$ in terms of radial velocity have been observed. Because bands of interstellar C$_{60}$ may be affected similarly, the $\gamma_0$ and A$_1$ bands may be shifted respectively up to 0.2~{\AA} and 0.13~{\AA} toward shorter or longer wavelengths. This is not a major issue considering the respective FWHMs of 1.5 $\pm$ 0.5~{\AA} and 5.54 $\pm$ 0.80~{\AA} that the two bands may show (Sect.~\ref{sec:labpwnt}).

The near-UV and violet regions show several lines of interstellar atoms and molecules that can serve as references, among which \ion{K}{I}, CN, CH, and CH$^+$ are found \citep[see, e.g.,][]{Hobbs09,Galazutdinov15}. The \ion{K}{I} lines, at 4044.136 and 4047.208~{\AA} rest wavelengths, are too weak to be seen in most of the present spectra though. In several cases, the CN, CH, and CH$^+$ lines show velocity components, which can be distinguished by fitting peak profiles onto them where they are not well separated. As interstellar wavelength references, we have chosen the lines indicating the same velocity as the strongest CH and CN components, that is, we assume C$_{60}$ is most likely to be found in the denser regions of diffuse clouds and in translucent clouds. Table~\ref{tbl:ISMmol} presents the wavelengths of the interstellar diatomic lines seen by the observer and the associated average radial velocities $\overline{v}_\mathrm{obs}$. They were determined by fitting Gaussian profiles to the diatomic lines as well as to the background stellar absorptions whenever it was necessary and possible.

\begin{table*}
\caption{Wavelengths and average radial velocities in the observer rest frame for interstellar molecules.}
\label{tbl:ISMmol}
\centering
\begin{tabular}{lllllllll}
\hline\hline
LOS & $\lambda_\mathrm{CN}$ ({\AA}) & $\lambda_\mathrm{CN}$ ({\AA}) & $\lambda_\mathrm{CN}$ ({\AA}) & $\lambda_\mathrm{CH}$ ({\AA}) & $\lambda_{\mathrm{CH}^+}$ ({\AA}) & $\lambda_{\mathrm{CH}^+}$ ({\AA}) & $\lambda_\mathrm{CH}$ ({\AA}) & $\overline{v}_\mathrm{obs}$ (km s$^{-1}$) \\
\hline
HD~76341  &   ...   & 3875.05 &   ...   & 3886.85 & 3958.14 & 4233.03 & 4300.81 & +33.9 \\
          &   ...   & (+0.44) &   ...   & (+0.44) & (+0.43) & (+0.48) & (+0.50) & \\
HD~136239 & 3873.51 & 3874.12 & 3875.27 & 3885.91 & 3957.20 & 4232.02 & 4299.77 & -38.1 \\
					& (-0.49) & (-0.49) & (-0.50) & (-0.50) & (-0.51) & (-0.53) & (-0.54) & \\
HD~148379 & 3873.37 & 3873.97 &   ...   & 3885.77 & 3957.06 & 4231.87 & 4299.61 & -49.0 \\
          & (-0.63) & (-0.64) &   ...   & (-0.64) & (-0.65) & (-0.68) & (-0.70) & \\
HD~167264 &   ...   &   ...   &   ...   & 3886.03 & 3957.30 & 4232.13 & 4299.89 & -29.9 \\
          &   ...   &   ...   &   ...   & (-0.38) & (-0.41) & (-0.42) & (-0.42) & \\
HD~168625 &   ...   & 3874.79 &   ...   & 3886.61 &   ...   &   ...   & 4300.54 & +15.1 \\
					&   ...   & (+0.18) &   ...   & (+0.20) &   ...   &   ...   & (+0.23) & \\
HD~169454 & 3874.12 & 3874.73 & 3875.89 & 3886.53 & 3957.81 & 4232.68 & 4300.46 & +9.2 \\
          & (+0.12) & (+0.12) & (+0.12) & (+0.12) & (+0.10) & (+0.13) & (+0.15) & \\
HD~183143 & 3874.29 & 3874.90 & 3876.06 & 3886.70 & 3957.98 & 4232.86 & 4300.63 & +22.1 \\
          & (+0.29) & (+0.29) & (+0.29) & (+0.29) & (+0.27) & (+0.31) & (+0.32) & \\
\hline
\end{tabular}
\tablefoot{Numbers between parentheses are shifts from rest values given by \citet{Hobbs09}. Three-$\sigma$ error smaller than 0.01~{\AA} for 75\% of the fitted wavelengths. One-$\sigma$ error greater than 0.01~{\AA} in one case only.}
\end{table*}


\section{Matrix-isolation spectroscopy of C$_{60}$ in solid Ne}\label{apx:MIS}

The equipment used for measuring the spectrum of C$_{60}$ isolated in Ne ice was described in previous studies \citep[e.g.,][]{Rouille09}. Conditions and parameters specific to the present case are as follows. Fullerene C$_{60}$ powder (SES Research, Inc., purity 99.5\%) was heated to 400~$^\circ$C and the molecules thus released through an opening in the oven were deposited together with an excess of Ne atoms (Air Liquide, purity 99.999\%) onto a 2~mm-thick CaF$_2$ substrate (Korth Kristalle GmbH) cooled to 7.5~K. The Ne gas was fed into the vacuum chamber at the mass rate of 5~sccm (standard cubic centimeter per minute) and the matrix deposition process lasted 50~min. Absorption spectra were derived from transmission measurements with a spectrophotometer that scanned wavelengths with a step of 0.2~nm, a resolution of 0.2~nm, and a rate of 11~nm~min$^{-1}$. Factory calibration of the wavelength scale was given as accurate to 0.3~nm.


\end{appendix}
\end{document}